\documentclass[aps,prd,showpacs,floatfix,nofootinbib,12pt]{revtex4}

\usepackage{color,amsmath,amssymb,graphicx,latexsym,subfigure}
\usepackage{threeparttable}
\usepackage[dvipsnames]{xcolor}
\usepackage{hyperref}

\newcommand{\sigsip}{\ensuremath{\sigma^{\rm{SI}}_p}}
\newcommand{\mchi}{\ensuremath{m_{\chi}}}

\newcommand{\gev}{\ensuremath{\,\mathrm{GeV}}}

\newcommand{\tev}{\ensuremath{\,\mathrm{TeV}}}
\newcommand{\kpc}{\ensuremath{\,\mathrm{kpc}}}

\usepackage{ulem,fancyvrb}
\usepackage{xcolor}

\usepackage{listings}
\usepackage{color}

\definecolor{dkgreen}{rgb}{0,0.6,0}
\definecolor{gray}{rgb}{0.5,0.5,0.5}
\definecolor{mauve}{rgb}{0.58,0,0.82}

\lstdefinestyle{BASHstyle}{frame=tb,
backgroundcolor=\color{gray!20},
  language=sh,
  aboveskip=3mm,
  belowskip=3mm,
  showstringspaces=false,
  columns=flexible,
  basicstyle={\scriptsize\ttfamily},
  numbers=none,
  numberstyle=\tiny\color{gray},
  keywordstyle=\color{blue},
  commentstyle=\color{dkgreen},
  stringstyle=\color{mauve},
  breaklines=true,
  breakatwhitespace=true,
  tabsize=3
}
\lstdefinestyle{F95style}{frame=tb,
backgroundcolor=\color{yellow!30},
  language=[95]Fortran,
  aboveskip=3mm,
  belowskip=2mm,
  showstringspaces=false,
  columns=flexible,
  basicstyle={\scriptsize\ttfamily},
  numbers=none,
  numberstyle=\tiny\color{gray},
  keywordstyle=\color{blue},
  commentstyle=\color{dkgreen},
  stringstyle=\color{mauve},
  breaklines=true,
  breakatwhitespace=true,
  tabsize=1
}

\lstnewenvironment{BASH}{\pagebreak[2]\lstset{style=BASHstyle}}{\pagebreak[2]}
\lstnewenvironment{F95}{\pagebreak[2]\lstset{style=F95style}}{\pagebreak[2]}

\begin{document}

{\small
\begin{flushright}
NCTS-PH/1717
\end{flushright} }

\voffset 1.25cm

\title{A combined analysis of PandaX, LUX, and XENON1T experiments   
within the framework of dark matter effective theory}

\author{Zuowei Liu$^1$\footnote{zuoweiliu@nju.edu.cn}}
\author{Yushan Su$^1$\footnote{151150045@smail.nju.edu.cn}}
\author{Yue-Lin Sming Tsai$^{1,2}$\footnote{sming.tsai@cts.nthu.edu.tw}}
\author{Bingrong Yu$^{1}$\footnote{bingrongyu123@gmail.com}}
\author{Qiang Yuan$^{3,4}$\footnote{yuanq@pmo.ac.cn}}
\affiliation{
$^1$School of Physics, Nanjing University, Nanjing, 210093, China\\
$^2$Physics Division, National Center for Theoretical Sciences, Hsinchu, Taiwan\\
$^3$Key Laboratory of Dark Matter and Space Astronomy, Purple Mountain
Observatory, Chinese Academy of Sciences, Nanjing 210008, China\\
$^4$School of Astronomy and Space Science, University of Science and 
Technology of China, Hefei 230026, China
}

\begin{abstract}
Weakly interacting massive particles are a widely well-probed dark matter 
candidate by the dark matter direct detection experiments. 
Theoretically, there are a large number of ultraviolet completed models 
that consist of a weakly interacting massive particle dark matter.  
The variety of models makes the comparison with the direct detection data 
complicated and often non-trivial. 
To overcome this, in the non-relativistic limit, the effective theory 
was developed in the literature which works 
very well to significantly reduce the complexity of dark matter-nucleon interactions 
and to better study the nuclear response functions.  
In the effective theory framework for a spin-1/2 dark matter, we 
combine three independent likelihood functions from the latest PandaX, 
LUX, and XENON1T data, and give a joint limit on each effective coupling. 
The astrophysical uncertainties of the dark matter distribution are also included 
in the likelihood. We further discuss the isospin violating cases of the
interactions. 
Finally, for both dimension-five and dimension-six effective theories 
above the electroweak scale, we give updated limits of the new physics mass scales.  
\end{abstract}
\date{\today}

%95.35.+d: Dark matter
%96.50.S-: Cosmic rays
%\pacs{95.35.+d,96.50.S-}

\maketitle

%%#######################################################%%
\section{Introduction \label{sec:intro}}
%%#######################################################%%

The search for particle dark matter (DM) is one of the most important 
topics in both particle physics and astrophysics. 
Yet, no clear evidence
appears in direct, indirect, or collider searches of the weakly interacting 
massive particles (WIMPs), which are the most intriguing DM candidates as
motivated by the thermal production of DM and its relic density. Instead,
several stringent limits have been reported, which pushes the WIMP mass 
heavier or the interaction between the DM and standard model (SM) particles 
weaker. Among those experiments, the direct detection ones have made
rapidly improved limits on probing the interaction between DM and the quark 
sector of SM. Such a limit from xenon-type detectors has been improved by
more than one order of magnitude in recent years, from XENON100 with 34 kg
target~\cite{Aprile:2012nq}, LUX with 250 kg~\cite{Akerib:2016vxi}, PandaX 
with 500 kg~\cite{Tan:2016zwf}, and finally to a ton-scale detector 
XENON1T~\cite{Aprile:2017iyp}. For a WIMP mass $m_\chi\sim 30\gev$, the 
latest XENON1T experiment sets upper limits on the spin-independent
WIMP-nucleon cross section to $\sim10^{-46}$ cm$^2$, which is only two 
orders of magnitude higher than the neutrino floor. 

These stringent limits constrain DM models severely. 
For some well-motivated DM models, the parameter space has significantly shrunk or 
the survival region needs to be fine-tuned, such as the blind-spot region for 
the neutralino DM in the supersymmetric models, (see e.g.  
Ref.~\cite{Choudhury:2015lha,Banerjee:2016hsk,Han:2016qtc} 
for the current status of the blind-spot region and 
Ref.~\cite{Athron:2017yua,Athron:2017qdc} for the latest comprehensive global study). 
On the other hand, the effective theory approach begins to catch more 
attention, such as the spin-1 mediator models 
including the Anapole, magnetic dipole, and electric dipole DM. 
The DM dipole interaction with the SM photon can be generated by a new mediator that 
is kinetically mixed with the SM photon. Such a mechanism can give rise 
to a velocity-dependent cross section~\cite{Gluscevic:2015sqa}.
In such a velocity-dependent framework, the published experimental limits for 
spin-independent and spin-dependent WIMP-nucleon cross section cannot be applied directly. 
Although there are many more WIMP candidate ultraviolet (UV) 
completed models than the given examples here, some of them may 
result in similar phenomena in direct detection experiments 
at low energies. Therefore, a model-independent limit 
from these experiments will be very useful to link the WIMP models with 
the direct detection experiments. 

Regardless of the model complexity, the DM direct detection is to search 
for recoil events due to the WIMP scattering off nuclei inside 
the detector. Because of the very small 
momentum transfer $ q \sim \mathcal{O}(\text{MeV})$ (for WIMPs) 
compared with mediator masses, such {an} interaction can be expressed 
using the effective field theory (EFT) whose heavier mediators can be 
integrated out and only a new physical scale $\Lambda$ and the information 
of spin and initial velocity are left. The low-energy EFT of DM has been 
extensively studied (e.g., see~\cite{Cao:2009uw,Goodman:2010yf,Fan:2010gt,Bai:2010hh,Goodman:2010ku,
Fox:2011fx,Fox:2011pm,Rajaraman:2011wf,Cheung:2012gi,
Matsumoto:2014rxa,Hoferichter:2015ipa,Hoferichter:2016nvd}). 
To study the EFT in a model independent framework, with possibly the spin 
and velocity dependence, a classification of 14 operators 
based on the spin and velocity of DM and target nuclei 
has been widely adopted in 
literature~\cite{DelNobile:2013sia,Fitzpatrick:2012ix,Anand:2013yka,
Gresham:2014vja,Catena:2014uqa,Catena:2014epa,Catena:2014hla,
Schneck:2015eqa,Catena:2015uua,Catena:2015vpa,Yang:2016wrl,
Bruggisser:2016nzw,Bishara:2016hek,Bishara:2017nnn,Witte:2016ydc,
Aprile:2017aas}.\footnote{The original study of operator 
analysis was Ref.~\cite{Dobrescu:2006au}, which was not in the context of DM.} 
The interaction between the DM and nuclei is generally expressed by 
their response functions~\cite{DelNobile:2013sia, Anand:2013yka}. 
Any UV completed model can then be described with a combination of these 
14 model-independent operators. This framework has also been adopted in 
studying the DM captured inside the Sun~\cite{Catena:2015uha,
Liang:2016yjf,Catena:2016ckl,Kavanagh:2016pyr,Catena:2016kro}.

In this work, we update the current limits on the EFT operators of DM
from the most recent PandaX, LUX, and XENON1T results, adopting similar methodology
as that of Ref.~\cite{Anand:2013yka}. Comparing with previous EFT works, 
we have the following three main improvements. Firstly, we reconstruct 
three likelihoods of PandaX, LUX, and XENON1T with proper consideration
of the astrophysical uncertainties. Secondly, using the reconstructed 
likelihood functions, we obtain combined limits on the DM mass and
coupling parameters for the 14 operators as given in Ref.~\cite{Anand:2013yka}. 
For a more generic purpose, we also discuss 
the situation of the relativistic Lagrangian which can consist of
more than one operator in the non-relativistic (NR) limit. We then discuss the 
isospin conserving (ISC) and isospin violating (ISV) scenarios. 
The ratios of maximum cancellation for ISV at the event rate level are 
computed. Finally, we show the updated lower bounds of new physics 
energy scale (or mediator mass) $\Lambda$ for dimension-five and 
dimension-six DM effective theories above electroweak symmetry breaking 
scale. The code and the likelihoods of experiments will be incorporated 
in the \texttt{LikeDM} tool~\cite{Huang:2016pxg}. 
An alternative package \texttt{GAMBIT}~\cite{Athron:2017ard} also solves the similar problems.

This paper is organized as follows. In Sec.~\ref{sec:direct} we briefly 
introduce the theory of DM direct detection in terms of effective operators. 
In Sec.~\ref{sec:likelihood}, we describe the experimental results of 
PandaX, LUX, and XENON1T, and their likelihood functions, including a 
detailed discussion of our treatment of the DM astrophysical parameters. 
In Sec.~\ref{sec:results} we present the results of our scans. 
We summarize our findings in Sec.~\ref{sec:sum}.

%%#######################################################%%
\section{DM direct detection theory\label{sec:direct}}
%%#######################################################%%
When the Earth sweeps through the local halo together with the sun, the 
DM from local halo may hit the underground target nuclei via the DM-nucleus scattering. 
The DM direct detection is designed to detect the 
nuclear recoil energy due to such an interaction. Unfortunately, no firm
detection of such interactions has been reported in present leading xenon-target experiments, 
such as PandaX-II~\cite{Tan:2016zwf}, LUX~\cite{Akerib:2016vxi}, and 
XENON1T~\cite{Aprile:2017iyp}. Usually the null result is interpreted as an upper
limit on the cross section of the DM-nucleon interaction, and presented 
separately on the spin-independent and spin-dependent components. 
To date, the most stringent limit on the spin-independent cross section 
$\sigsip$ comes from the XENON1T experiment \cite{Aprile:2017iyp}.

\begin{table}[t]
    \centering
    \begin{tabular}{ll}
    \hline\hline
        %\zx{${\mathcal{O}}_1 = \textbf{1}_{\chi N}$}
         {${\mathcal{O}}_1 = \textbf{1}_{\chi}\textbf{1}_{N}$}
         & ${\mathcal{O}}_9 = i{\bf{{S}}}_\chi\cdot\left({{\bf{S}}}_N\times\frac{{\bf{{q}}}}{m_N}\right)$  \\
        ${\mathcal{O}}_3 = i{{\bf{S}}}_N\cdot\left(\frac{{\bf{{q}}}}{m_N}\times{\bf{{v}}}^{\perp}\right)$ \hspace{2 cm} &   ${\mathcal{O}}_{10} = i{{\bf{S}}}_N\cdot\frac{{\bf{{q}}}}{m_N}$   \\
        ${\mathcal{O}}_4 = {{\bf{S}}}_{\chi}\cdot {{\bf{S}}}_{N}$ &   ${\mathcal{O}}_{11} = i{\bf{{S}}}_\chi\cdot\frac{{\bf{{q}}}}{m_N}$   \\                                                                             
        ${\mathcal{O}}_5 = i{\bf{{S}}}_\chi\cdot\left(\frac{{\bf{{q}}}}{m_N}\times{\bf{{v}}}^{\perp}\right)$ &  ${\mathcal{O}}_{12} = {{\bf{S}}}_{\chi}\cdot \left({{\bf{S}}}_{N} \times{\bf{{v}}}^{\perp} \right)$ \\                                                                                                                 
        ${\mathcal{O}}_6 = \left({\bf{{S}}}_\chi\cdot\frac{{\bf{{q}}}}{m_N}\right) \left({{\bf{S}}}_N\cdot\frac{{{\bf{q}}}}{m_N}\right)$ &  ${\mathcal{O}}_{13} =i \left({{\bf{S}}}_{\chi}\cdot {\bf{{v}}}^{\perp}\right)\left({{\bf{S}}}_{N}\cdot \frac{{\bf{{q}}}}{m_N}\right)$ \\   
        ${\mathcal{O}}_7 = {{\bf{S}}}_{N}\cdot {\bf{{v}}}^{\perp}$ &  ${\mathcal{O}}_{14} = i\left({{\bf{S}}}_{\chi}\cdot \frac{{\bf{{q}}}}{m_N}\right)\left({{\bf{S}}}_{N}\cdot {\bf{{v}}}^{\perp}\right)$  \\
        ${\mathcal{O}}_8 = {{\bf{S}}}_{\chi}\cdot {\bf{{v}}}^{\perp}$  & ${\mathcal{O}}_{15} = -\left({{\bf{S}}}_{\chi}\cdot \frac{{\bf{{q}}}}{m_N}\right)\left[ \left({{\bf{S}}}_{N}\times {\bf{{v}}}^{\perp} \right) \cdot \frac{{\bf{{q}}}}{m_N}\right] $ \\                                                                               
    \hline\hline
    \end{tabular}
    \caption{Non-relativistic quantum mechanical operators defining the 
general effective theory of one-body DM-nucleon interactions. Taken from 
Ref.~\citep{Anand:2013yka}.} 
    \label{tab:operators}
\end{table}

Generally the exclusion limits by the experiment groups are based on 
simplified assumptions and cannot be used directly for 
a number of interactions. 
For example, a possible enhancement at small momentum transfer for 
the velocity dependent scattering cross section is usually not properly investigated in the 
standard spin-independent {or} spin-dependent approach by the experimental groups.  
To study the constraints on a variety of different operators,
we adopted the particle model independent method developed in 
Refs.~\citep{Fitzpatrick:2012ix,Anand:2013yka}. The non-relativistic 
quantum mechanical operators are listed in the Table~\ref{tab:operators}. 
The operators $\mathcal{O}_1$ to $\mathcal{O}_{11}$ are associated with 
interactions of spin-0 or spin-1 mediators, while the rest are not.
Because the operators in Table~\ref{tab:operators} are the mass dimension six operators, 
it is convenient to introduce some mass-scale parameters into the coefficients 
to make {them} dimensionless.
Hence, by including the physical scale $\Lambda$ and expanding the 
isospin index of coefficients, one can rewrite the operator as
\begin{equation}
\left [ \frac{c_i^p}{\Lambda^2}\frac{1+\tau_3}{2}+ 
\frac{c_i^n}{\Lambda^2}\frac{1-\tau_3}{2}  \right ]\mathcal{O}_i, 
\end{equation}
where $\tau_3$ is the 3rd Pauli matrix and 
%\zx{$i$ denotes the \zx{15}{14} different operators.}
{$i$ runs from 1 to 15 for different operators with ${\cal O}_2$ omitted.}
Following the convention in the code 
\texttt{DMFormFactor}~\citep{Anand:2013yka}, 
we set $\Lambda$ to be the Higgs vacuum expectation value $\Lambda=246\gev$. 
The new coefficients $c_i^p$ and $c_i^n$ are treated as input parameters, 
and the relations $c_i^p=c_i^n$ and $c_i^p\ne c_i^n$ correspond to ISC 
and ISV scenarios, respectively.

In the limits of small recoil energy and slow DM velocity, one can expand 
all possible effective interactions with the following four three-vectors in the non-relativistic limit, 
\begin{equation}
{{\bf{S}}}_{\chi},\,{{\bf{S}}}_{N},\,i\frac{{\bf{{q}}}}{m_N},\,\,{\bf{{v}}}^{\perp},
\label{Eq:fourvectors}
\end{equation}
where ${{\bf{S}}}_{\chi}$ and ${{\bf{S}}}_{N}$ are spins of the DM and 
nucleon, the vector $\bf{{q}}$ is the transfer momentum, and 
${\bf{{v}}}^{\perp}$ is the velocity that is perpendicular to the momentum transfer $\bf{{q}}$.
The velocity ${\bf{{v}}}^{\perp}$ is defined as ${\bf{{v}}}^{\perp}={\bf v}+{\bf q}/(2\mu_{\chi N})$ where 
$\mu_{\chi N}$ is the dark matter-nucleon reduced mass.
%\begin{equation}
%{\bf{{v}}}^{\perp}=\frac{1}{2}\left(
%{v}_{\chi,\rm in}+{v}_{\chi,\rm out}-{v}_{N,\rm in}-{v}_{N,\rm out}
%\right).
%\label{Eq:perpvel}
%\end{equation}
In classical limits, these four three-vectors describe the states before and 
after a scattering. Generally speaking, if the operators are nucleon 
spin-dependent, the vector ${\bf{S}}_{N}$ appears in the operators whose 
cross sections are not coherently enhanced by a factor of target atom 
number square as the nucleon spin-independent case. Regarding the 
velocity dependent operators, they may come from the anapole or dipole 
interaction. For example, in Ref.~\cite{Witte:2016ydc}, it is shown
that the anapole interaction $\bar{\chi}\gamma_5\chi$ depends on 
$\bf{q}^2$; the magnetic dipole interaction $\bar{\chi}\sigma^{\mu\nu}\chi$ 
depends on $\bf{q}^4$ and $(\bf{q}{\bf{v}}^{\perp})^2$ ($({\bf{v}}^{\perp}/
\bf{q})^2$) for the case of mediator mass heavier (lighter) than the 
momentum transfer; the electric dipole interaction 
$\bar{\chi}\sigma^{\mu\nu}\gamma_5\chi$ depends on $\bf{q}^2$ ($\bf{q}^{-2}$) 
for the case of mediator mass heavier (lighter) than the momentum transfer.

Following the conventions of Ref.~\cite{Catena:2015uua}, we write the 
differential event rate of scattering between DM and the target nuclues
\textit{per unit detector mass} as a function of the recoil energy $Q$ as  
\begin{eqnarray}
\frac{{\rm d}\mathcal{R}}{{\rm d}Q} &=& \sum_{T} \xi_T \frac{\rho_{0}}{m_\chi m_{T} }  
 \int_{v > v_{\rm min}(Q)} \,  v  f(\vec{v} + \vec{v}_e)\frac{{\rm d}\sigma}{{\rm d}Q}\, d^3v,
\label{Eq:dndQ}
\end{eqnarray}
where $m_\chi$ and $m_T$ are the DM and target masses. The parameter $\xi_T$ 
is defined as 
\begin{equation}
\xi_T =  \frac{\eta_{T} m_T}{\sum\limits_{T} \eta_{T} m_T}, 
\label{Eq:xiT}
\end{equation}
where $\eta_T$ can be found in the website~\footnote{\url{https://www.webelements.com/xenon/}}. 
The differential cross section is
\begin{eqnarray}
\frac{{\rm d}\sigma}{{\rm d}Q} &=& \frac{m_{T}}{2\pi v^2}
\langle |\mathcal{M}_{NR}|^2\rangle_{\rm spins},
%ProbDen(j)*mT/(2d0*pi*v(j)**2)
%\frac{{\rm d}\mathcal{R}}{{\rm d}Q} =  \sum_{T} \xi_T \frac{\rho_{0}}{2\pi m_\chi}  
% \int_{v > v_{\rm min}(Q)} \,  \frac{f(\vec{v} + \vec{v}_e)}{v} \, 
% \langle |\mathcal{M}_{NR}|^2\rangle_{\rm spins} \, d^3v, 
\label{Eq:dXSdQ}
\end{eqnarray}
and the averaged amplitude can be written as  
\begin{align}
\langle |\mathcal{M}_{NR}|^2\rangle_{\rm spins} =  
\frac{4\pi}{2J+1}\sum_{\tau,\tau'=\{p,n\}} 
&
\bigg[ \sum_{k=\{M,\Sigma',\Sigma''\}} R^{\tau\tau'}_k\left(v_T^{\perp 2}, {q^2 \over m_N^2} \right) W_k^{\tau\tau'}(y) \nonumber\\
&
+{q^{2} \over m_N^2} \sum_{k=\{\Phi'', \Phi'' M, \tilde{\Phi}', 
\Delta, \Delta \Sigma'\}} R^{\tau\tau'}_k\left(v_T^{\perp 2}, {q^2 \over m_N^2}\right) W_k^{\tau\tau'}(y) \bigg] 
\,. \nonumber\\
\label{eq:amplitudes}
\end{align}
The amplitude includes the nuclear response 
functions $W_k^{\tau\tau'}(y)$ and the DM response functions $R^{\tau\tau'}_k$,
which were described in Ref.~\cite{Anand:2013yka}. 
The indices $\tau$ and $\tau'$ run over proton $p$ and neutron $n$.
We also briefly describe
them in Appendix~\ref{sec:DM_respond}. 
The symbols $M$, $\Delta$, $\Sigma^\prime$, $\Sigma^{\prime\prime}$, 
$\tilde{\Phi}^\prime$, and $\Phi^{\prime\prime}$ indicate the type of response function. 
Note that the dimension of the averaged amplitude is $\rm{mass}^{-4}$.
One shall not confuse the transfer momentum $q$ in Eq.~\eqref{eq:amplitudes} 
with the recoil energy $Q$ in Eq.~\eqref{Eq:dXSdQ}. These two quantities
are related by $q^2=2 m_T Q$.

In the astrophysics part, the DM velocity distribution function $f(\vec{v} 
+ \vec{v}_e)$ can be described by the Maxwell-Boltzmann
distribution~\cite{Lisanti:2010qx}, or directly extracted from $N$-body 
simulation~\cite{Bozorgnia:2016ogo}. The Earth's velocity in the galactic 
rest frame, $\vec{v}_e$, is added in order to translate the reference frame 
from the galactic rest frame to the Earth rest frame. In this work, we adopt 
the soft-truncated Maxwell-Boltzmann distribution~\cite{Savage:2008er}. 
The local density $\rho_{0}$ indicates the DM mass density near the Sun.
We analytically integrate 
the velocity distribution over the solid angle and present the result in 
Appendix~\ref{sec:DM_vf}. The minimum DM velocity can be written as a 
function of recoil energy, assuming that the scattering between DM and 
nucleus is elastic
\begin{equation}
v_{\rm{min}}(Q)=\sqrt{\frac{Q m_T}{2M_r^2}},
\end{equation}
where $M_r$ is the reduced mass. 
 
Finally, the efficiency of {the dark matter} detector $\epsilon(Q)$ needs to be included 
in the calculation. 
The total event rate is then
\begin{equation}
\mathcal{R}=\int^{\infty}_{0}\epsilon(Q)\frac{{\rm d}\mathcal{R}}{{\rm d}Q} dQ.
\label{Eq:events}
\end{equation} 
We will see in the following section that the event rate $\mathcal{R}$ is 
the best model independent quantity in our likelihood functions. 
 
%To compute the time averaged number of events predicted by the theory,
%one has first consider the energy resolution, $\kappa(Q,E)$, to convolve the case that
%a recoil energy $Q$ is measured as $E$.
%In addition, the the efficiency of detector, $\epsilon(E)$,
%has to be included, so that the total time averaged number of events read as
%\begin{equation}
%N=\int^{\infty}_{0} \epsilon(E) dE \int^{\infty}_{0} \kappa(Q,E)\frac{{\rm d}\mathcal{R}}{{\rm d}Q} dQ.
%\label{Eq:events}
%\end{equation} 

\section{Experimental Data and Likelihood function\label{sec:likelihood}}

\subsection{PandaX-II experiment}
The PandaX-II is a half-ton xenon dual-phase detector at the China 
Jinping underground Laboratory. The PandaX collaboration has published 
results based on their Run 8 (19.1 live days) and Run 9 (79.6 live days) 
data with exposure of 5845 kg-day and 27155 kg-day, respectively. For the Run 8, 
the total observed event number after all the cuts is 2 
%\zx{while expected}
and the expected background event number is
$2.4\pm0.8$. For the Run 9, the observed event number is 1 with an expected 
background event number
of $2.4\pm0.7$. The $90\%$ confidence level (CL) upper limit 
on the $(m_\chi,\sigsip)$ 
panel has been shown in Fig.~5 of Ref.~\cite{Tan:2016zwf}. 
Usually, the $90\%$ CL limit in the experimental results of the DM direct detection, 
is computed with the two-tail convention, which corresponds 
to the $95\%$ CL limit in the one-tail convention. 
We will adopt the one-tail convention in the analysis.

%In reality, it is difficult to exactly reproduce the above results due
%to limited information about the experiment details. 
%However, {\color{blue}{it is still possible to do some tricks with this 
%difference.}}
%In this work, we adopt the method developed in Ref.~\cite{Cheung:2012xb} 
%to build the likelihood functions.
     
Based on the data of PandaX-II, one can simply build the likelihood 
function as~\cite{Cheung:2012xb}
\begin{equation}
\mathcal{L}_{\rm{PandaX}}\propto \prod_{i=\rm{run8,run9}} 
\max_{b_i^\prime}\frac{\exp[-(s_i+b_i^\prime)] (s_i+b_i^\prime)^{o_i}}{o_i!} 
\exp\left[-\frac{(b_i^\prime-b_i)^2}{2 \delta b_i^2}\right], 
\label{Eq:pandaxlike}
\end{equation}
where $b_8=b_9=2.4$, $\delta b_8=0.8$, $\delta b_9=0.7$, 
$o_8=2$, $o_9=1$. 
Here $s_i={\cal E}_i {\cal R}_\text{PandaX}$ is the theoretically predicted number of events, 
where ${\cal E}_8=5845$ kg-day, ${\cal E}_9=27155$ kg-day, 
and ${\cal R}_\text{PandaX}$ is computed via Eq.~\eqref{Eq:events}. 
The nuisance parameter $b_i^\prime$ is introduced for the running of
pseudo-experiments. 
From Eq.~\eqref{Eq:pandaxlike}, a DM model-independent likelihood function 
can be obtained for the PandaX experiment. 
Based on this pure statistical likelihood 
function with the $CL_b$ method (see Appendix ~\ref{sec:CLb}), 
the $95\%$ CL upper limit on the event rate ${\cal R}_\text{PandaX}$ 
is $7.11\times 10^{-5}$ per kg-day.

Due to the complexity of the experimental performance 
one usually cannot exactly reproduce the $\sigma^{\rm{SI}}_{p,95\%}$ line given in 
Ref.~\cite{Tan:2016zwf}. Therefore, we introduce an additional correction factor 
$f(m_\chi)$ to account for this discrepancy,
\begin{equation}\label{eq:astroR}
f(m_\chi) \equiv \frac{s_{95}^{\rm exp}(m_{\chi})}
                 {s_{95}^{\rm stat}}
\end{equation}
where $s_{95}^{\rm exp}(m_{\chi})$ is 
the event number inferred from the experimental limits, the 
$\sigma^{\rm{SI}}_{p,95\%}$ line from Ref.~\cite{Tan:2016zwf}, 
and $s_{95}^{\rm stat}$ is the $95\%$ CL limit computed 
via our likelihood analysis. 
The default
values for astrophysical parameters, quoted by Ref.~\cite{Tan:2016zwf},
are $v_0=220\,\rm{km}/s$, $\rho_0=0.3\,\gev/\rm{cm}^3$, and 
$v_{esc}=544\,\rm{km}/s$. $s_{95}^{\rm stat}$ is computed by 
Eq.~\eqref{Eq:pandaxlike}, which is DM model independent.
Taking into account the correction factor $f(m_\chi)$, we have 
\begin{equation}\label{eq:si}
s_{i}(m_\chi,\mathcal{R})=\frac{\mathcal{R E}_i}{f(m_{\chi})}, 
\end{equation}
which is then used 
to compute the likelihood function given by Eq.~\eqref{Eq:pandaxlike}, 
instead of $s_i =  {\cal RE}_i$ as done previously. 
With this correction factor included in the likelihood function, 
we are able to reproduce the PandaX results. Thus, in our analysis, 
the total likelihood of PandaX-II, 
$\mathcal{L}_{\rm{PandaX}}$, is a function of both $\mathcal{R}$ 
and $m_\chi$.

%If we consider the relationship between the event number and rate, we have
%\begin{equation}
%s_{i modify}(m_\chi,\mathcal{R})=R_{modify}(m_\chi,\mathcal{R})\times{\rm (exposure)_i}
%\end{equation}
%where
%\begin{equation}
%R_{modify}(m_\chi,\mathcal{R})=\mathcal{R}/f(m_{\chi})=\mathcal R\times\frac{R_{95}^{likelihood}}{R_{95}%^{theoretical}(m_{\chi})}
%\end{equation}

\subsection{LUX experiment}

The LUX experiment is also a dual-phase xenon (250 kg) time projection chamber. 
Combining two data sets \texttt{WS2013} and \texttt{WS2014-16},   
the LUX has a similar exposure ($3.35\times 10^{4}$ kg-day) to that of
the PandaX, and gives comparable but slightly stronger limits (perhaps due to a
different analysis method\footnote{The PandaX collaboration adopted the 
boosted-decision-tree (BDT) method to optimize the rejection of the 
nuclear recoil background, while the LUX group did not use the BDT. 
Furthermore, a time-dependent mapping between the true recoil position
and the observed S2 coordinate is required to interpret the LUX data. 
}), 
on the DM-nucleon cross section. 
The LUX experiment did not report the total observed 
and expected events after cuts, which makes the
construction of the likelihood function much more difficult. 
We construct a different likelihood function here than that in the PandaX case.
By assuming that
null signal has been detected, we build the total likelihood function 
of LUX as 
\begin{eqnarray}
&&\ln \mathcal{L}_{\rm{LUX}}=\sum_{i=\texttt{WS2013},\texttt{WS2014-16}}
 \ln \mathcal{L}_{i}(m_\chi,\mathcal{R}),  \nonumber \\
&&-2\ln \mathcal{L}_{i}(m_\chi,\mathcal{R})= 
\left[\frac{s_i(\mathcal{R})}{ s_{i,{\rm{95}}}(m_\chi)/1.64}\right]^2,
\label{Eq:luxlike}
\end{eqnarray}
where $s_{i,{\rm{95}}}(m_\chi)$ is the number of events computed from the 
$95\%$ CL limit curve in 
Ref.~\cite{Akerib:2016vxi}. 
Here, the $95\%$ CL is equivalent to $1.64\sigma$ far from the central value 
in 1-dimensional Gaussian likelihood.
One has to bear in mind that the 
theoretical prediction of the event number $s_i$ depends on the specific
particle model of DM, because the efficiencies of \texttt{WS2013} and 
\texttt{WS2014-16} are different and it is hard to compute the likelihood 
in advance with a particle model-independent quantity. Hence, for the 
purpose of a particle model-independent study, we employ two independent 
$\mathcal{R}$ variables and two likelihood functions. Since the result
of $\sigma^{\rm{SI}}_{p,95\%}$ of Ref.~\cite{Akerib:2016vxi} has been
used in the likelihood Eq.~\eqref{Eq:luxlike}, we do not need to consider 
the correction factor $f(m_{\chi})$ as in the PandaX likelihood.

\subsection{XENON1T Experiment}
The XENON1T~\cite{Aprile:2017iyp} is the first ton-scale xenon-type detector. 
With a fiducial mass of $\sim 1042$ kg and a running of 34.2 live days, the 
XENON1T data give the currently most stringent limit on the spin-independent 
cross section between DM and nucleon. Like PandaX and LUX, the XENON1T is 
also a dual-phase detector, and all the setups of the PandaX likelihood can 
be directly applied for XENON1T, with only the replacement of $o=1$
and $b=0.36^{+0.11}_{-0.07}$~\cite{Aprile:2017iyp}. 
A piecewise function is used to incorporate both the positive and the negative background uncertainties. 
The efficiency can be found in Fig. 1 of Ref.~\cite{Aprile:2017iyp}. Similarly, a correction 
factor $f(m_\chi)$, which is different from the one of PandaX, is needed 
to compensate the discrepancy between the official Xenon1T limit and 
our analysis.      

%Note that the observed event $o_i=1$ is read from the number of events 
%below the $95\%$ CL expected nuclear recoil line presented in the 
%Fig. 2c of Ref.~\cite{Aprile:2017iyp}. 
One has to bear in mind that both XENON1T and PandaX groups used the 
unbinned likelihood analysis. 
%it means this the observed event number 
%does not enter the likelihood functions explicitly. 
On the contrary, our likelihood is a binned method which relies on the 
event number. In the binned method, all the information of each event are folded 
which makes our constraints weaker. Hence, our likelihood is more conservative than the 
actual ones by XENON1T and PandaX data.

\subsection{Combined Likelihood}
Before combining three experimental data sets, it is necessary to consider 
the shared systematical uncertainties which may be multiply counted. 
Since LUX, PandaX, and XENON1T are three independent experiments, the 
systematical uncertainties from instrumentation are expected to be
independent. However, the astrophysical uncertainties of the DM 
distribution should be the same. They should only be considered once
in the likelihood calculation.
%All three experiments fix the astrophysical {\color{blue}{parameters}} 
%for both DM source and propagation so that one can ignore this part of 
%uncertainties and reintroduce it later. 
%In addition, our likelihood is built based on the event rate $\mathcal{R}$ 
%but not total events. Hence, it is safely to combine the likelihood from 
%LUX, PandaX, and XENON1T with equal weight to increase the statistics.

\begin{figure}[!htb]
\includegraphics[width=0.45\textwidth]{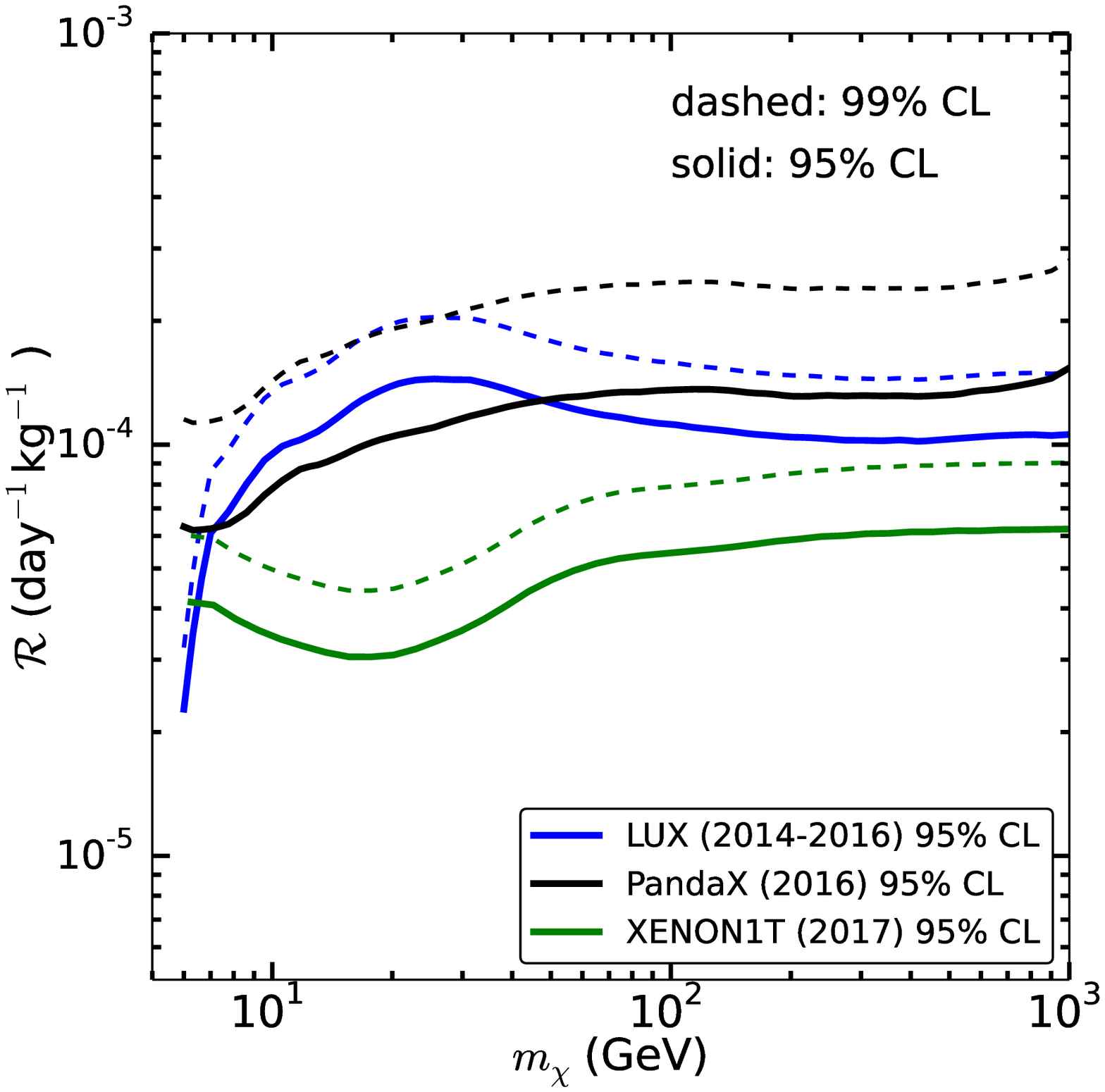}
\includegraphics[width=0.45\textwidth]{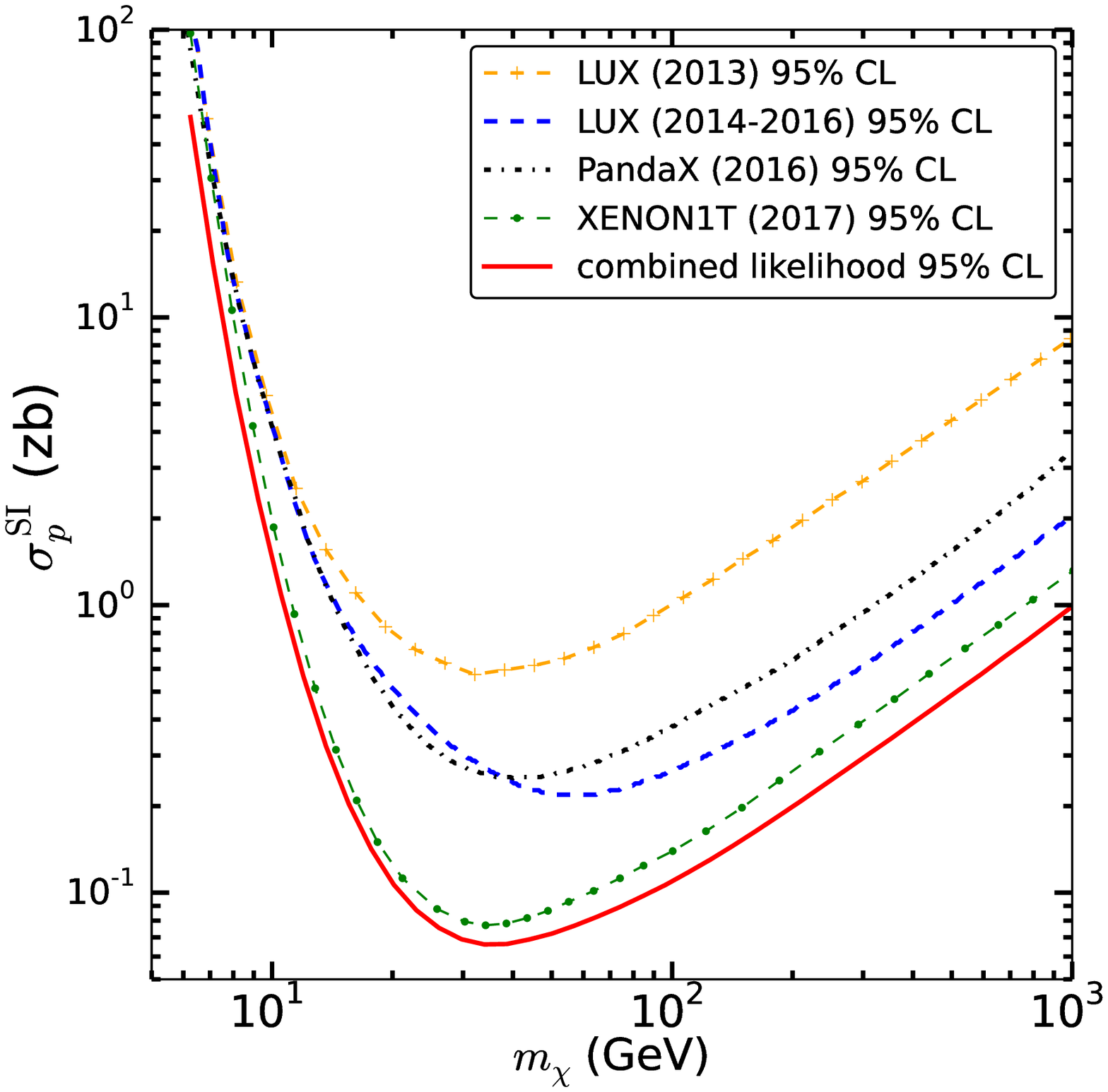}
\caption{Left panel: upper limits of the event rate $\mathcal{R}$ as a 
function of $m_\chi$ at $95\%$ (solid) and $99\%$ (dashed) CL, for PandaX 
(black), LUX (blue), and XENON1T (green). 
Right panel: the $95\%$ CL upper limits of $\sigsip$ for PandaX, LUX, 
XENON1T, and the combination of these three experiments. 
Here, we adopted exactly the same astrophysical setup as used 
in the experimental computation,  
the soft-truncated Maxwell-Boltzmann distribution with 
$v_0=220\,\rm{km}/s$, $\rho_0=0.3\,\gev/\rm{cm}^3$, and 
$v_\text{esc}=544\,\rm{km}/s$. 
\label{fig:CL_and_Comb}}
\end{figure}

The combined likelihood can be simply written as the product of 
likelihoods of the three experiments, PandaX-II \texttt{run8+run9}
\footnote{Note that the \texttt{run10} data~\cite{Cui:2017nnn} 
was published one week after our paper appeared in the arxiv. 
We also update the result in the appendix~\ref{sec:PX10}, 
but the new combined limit is only slightly improved.}, 
LUX \texttt{WS2013+WS2014-16}, and XENON1T. 
In the left panel of Fig.~\ref{fig:CL_and_Comb}, we show the upper 
limits of $\mathcal{R}$ at the $95\%$ (solid) and $99\%$ (dashed) CL for 
PandaX, LUX, and XENON1T. 
In all the mass region, the XENON1T limits
are lower than the other two experiments. The capabilities of PandaX 
and LUX in constraining the DM-nucleon cross section 
are similar with each other, with slightly different sensitive
mass regions. By comparing between the $95\%$ CL (solid) 
and $99\%$ CL (dashed) lines, 
we find that the PandaX has larger gap between the two lines than LUX, 
which can explain why LUX can constrain the parameter 
space more stringent than PandaX, although their exposure is comparable. 
In the right panel of Fig.~\ref{fig:CL_and_Comb}, 
we give the $95\%$ CL upper limits of the spin-independent cross section 
based on individual likelihood and the combined one. Our combined 
constraint improves by a factor of $\sim 1.3$ compared with the
XENON1T result.

\subsection{DM astrophysical nuisance parameters}
\label{sec:astro}
\begin{table}
%    !--- the value is taken from Table 4 of paper 1604.01216 ---
\begin{center}
\begin{tabular}{ll}
\hline
\hline
%\multicolumn{3}{|c|}{Astrophysical DM halo parameters} \\
%\hline\hline
$\rho_{0}$ ($ \gev~\rm{cm}^{-3}$) & $ 0.32\pm 0.02$ \\
$v_{0}$ (km~s$^{-1}$)& $240.0\pm 6.0$  \\
$v_{\rm esc}$ (km~s$^{-1}$)& $ 541.3^{+16.6}_{-12.2}$ \\
\hline
\hline
\end{tabular}
\end{center}
\caption{The DM astrophysical nuisance parameters given in 
Ref.~\cite{Huang:2016} which considers the rotation curve of the Milky Way 
within $\sim100\kpc$.
}
\label{tab:astro_params}
\end{table}

As aforementioned, the astrophysical uncertainties have to be properly 
considered, particularly for the operators where the cross section is 
velocity dependent. The DM velocity can be simply described by a soft 
truncated Maxwell-Boltzmann distribution (see Eq.~\eqref{eq:I2}). 
The parameters of the velocity distribution, the DM local velocity, 
the escape velocity, and the local density can be determined by the 
kinematics of stars or gas in the Milky Way~\cite{Huang:2016}.
%The DM $N$-body simulations are developed for such issues but it still
%remains a large systematical uncertainties due to the lacks of knowledge
%of the baryonic ingredients. On the other hand, if ignoring such unclear
%uncertainties, one can fit the data from primary red clump giants in the 
%outer disk assuming NFW halo profile, for example see Ref.~\cite{Huang:2016}. 
%The NFW halo profile is a fitting formula based on the result from 
%$\Lambda$CDM $N$-body simulation. 
Such methods are, however, subject to systematic uncertainties of the DM halo
profile. See Ref.~\cite{Read:2014qva} for a compilation of the results
on the local density measurements by different methods and/or data sets.

In this work we adopt the velocity parameters derived in 
Ref.~\cite{Huang:2016}, which employed an updated precise measurement 
of the rotation curve using the LAMOST data to determine the DM halo
properties. In Table~\ref{tab:astro_params}, we list the parameter values
and their $1\sigma$ uncertainties of the local DM density, the local DM
velocity, and the escape velocity given in Ref.~\cite{Huang:2016}. 
Unlike \texttt{GAMBIT DDCalc}~\cite{Workgroup:2017lvb} whose nuisance distribution of DM local density 
is log-normal, we adopt normal distribution.
The uncertainties of the local velocity and escape velocity will be
included in the likelihood calculation, via a profile likelihood
approach. 

Note that the integration of the DM velocity distribution
over the solid angle is computationally heavy; we thus provide three
analytical formulae valid in different velocity regions in 
Appendix~\ref{sec:DM_vf}. The local DM density affects the normalization
of the event rate. We also quote the results given in Ref.~\cite{Read:2014qva},
i.e., $0.20-0.56$ GeV cm$^{-3}$, which is indicated by a band in the plots of our final results.

\section{Numerical results \label{sec:results}}

In this section, the combined limits from the PandaX, LUX and XENON1T 
are presented on (i) the non-relativistic operator coefficients, 
(ii) the relativistic effective operator couplings, and (iii) the high energy 
EFT energy scale $\Lambda$. We only focus on the DM masses between 
$5\gev$ and $1\tev$. Note that our likelihood functions are 
particle model independent but only the spin-1/2 DM is considered in this work as illustration. 
For DM with different spins, the limits can be different.

We present the limits at the $95\%$ CL with the $CL_b$ hypothesis, 
namely $-2\Delta\ln\mathcal{L}=-2(\ln\mathcal{L}-\ln\mathcal{L}_0)=2.71$. 
%As argued in previous sections, our likelihood reconstruction is 
%not able to perfectly reproduce the experimental likelihoods but 
%attempts to mimic their likelihood tails.  
Here $\mathcal{L}_0$ is the likelihood without DM signal (background
only hypothesis), and $\mathcal{L}$ is the corresponding likelihood
for given DM model parameters. 
$\mathcal{L}_0$ is slightly
smaller than the global maximal
likelihood in the background-and-DM hypothesis, due to a less-than-$1\sigma$ excess for the XENON1T data.
The choice of the background-only hypothesis instead of the global
maximum $\mathcal{L}$ as reference makes our results more conservative.

The astrophysical parameters, including the local velocity and escape
velocity, are treated as nuisance parameters following Gaussian 
distributions with parameters given in Table~\ref{tab:astro_params}.
As we have discussed above, a larger uncertainty band of the local density
measurements ~\cite{Read:2014qva} is adopted to account for possible 
systematic uncertainties.

In many ISV studies~\cite{Feng:2011vu,Gao:2011ka,Kumar:2011dr,
Jin:2012jn,Okada:2013cba,Belanger:2013tla,Cirigliano:2013zta,
Hamaguchi:2014pja,Zheng:2014nga,Chen:2014tka,Martin-Lozano:2015vva,
Yaguna:2016bga}, the ISV coupling ratio for maximum cancellation 
between the DM-proton contribution and the DM-neutron contribution 
is defined at the amplitude level. Unlike such traditional definitions, 
we define our ISV coupling ratio for maximum cancellation at the 
event rate level, 
%to be the ratio $c_n/c_p$ located at $95\%$ C.L. lines 
in which we take the experimental efficiencies and the DM velocity 
distribution into account. This new definition shall be more generic. 
As a comparison, the ISC scenario ($c_n=c_p$) is 
also presented. 

%In this section, we will first show the limits at three different scales.
%{\color{blue}{Firstly}}, the $95\%$ C.L. for each sole non-relativistic operator coefficient {\color{blue}{will be shown}}. 
%{\color{blue}{Secondly}}, we will discuss the upper limits for relativistic effective Lagrangian couplings.          
%Finally, a lower limit of $\Lambda$ for higher energy theory is computed.   
           
\subsection{Non-relativistic operators}

\begin{figure}[!htb]
\includegraphics[width=0.4\textwidth]{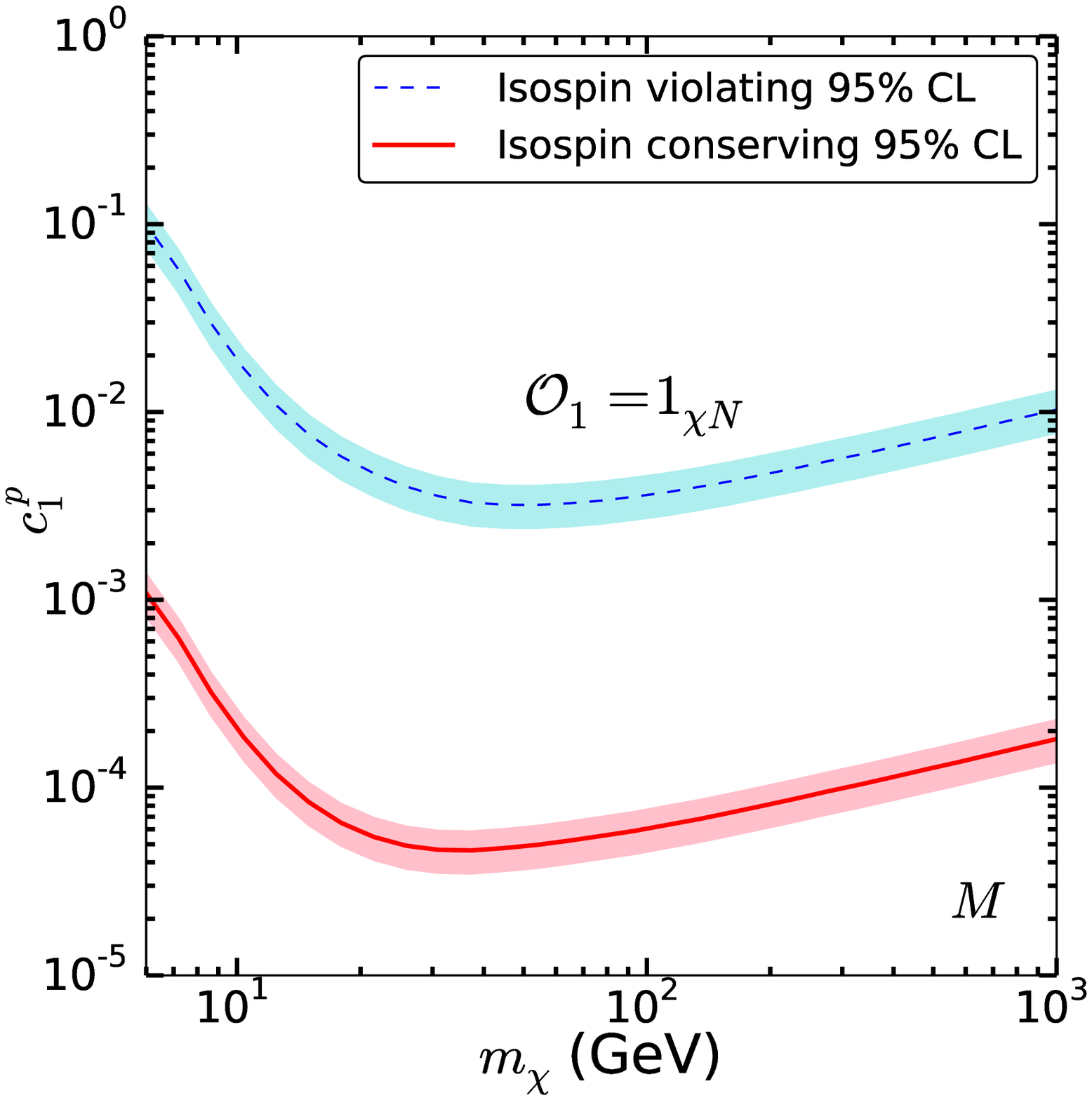}
\includegraphics[width=0.4\textwidth]{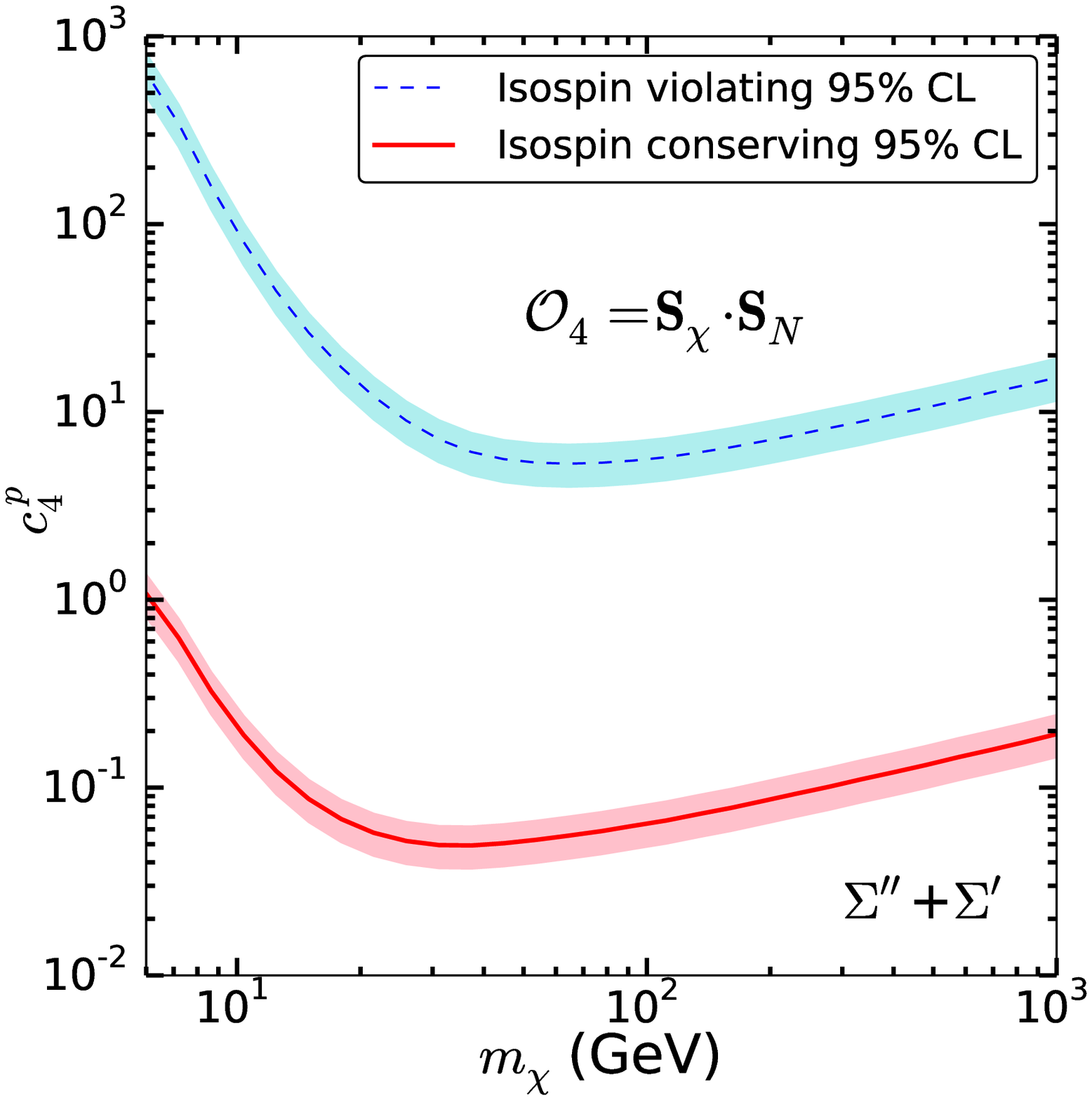}\\
\includegraphics[width=0.4\textwidth]{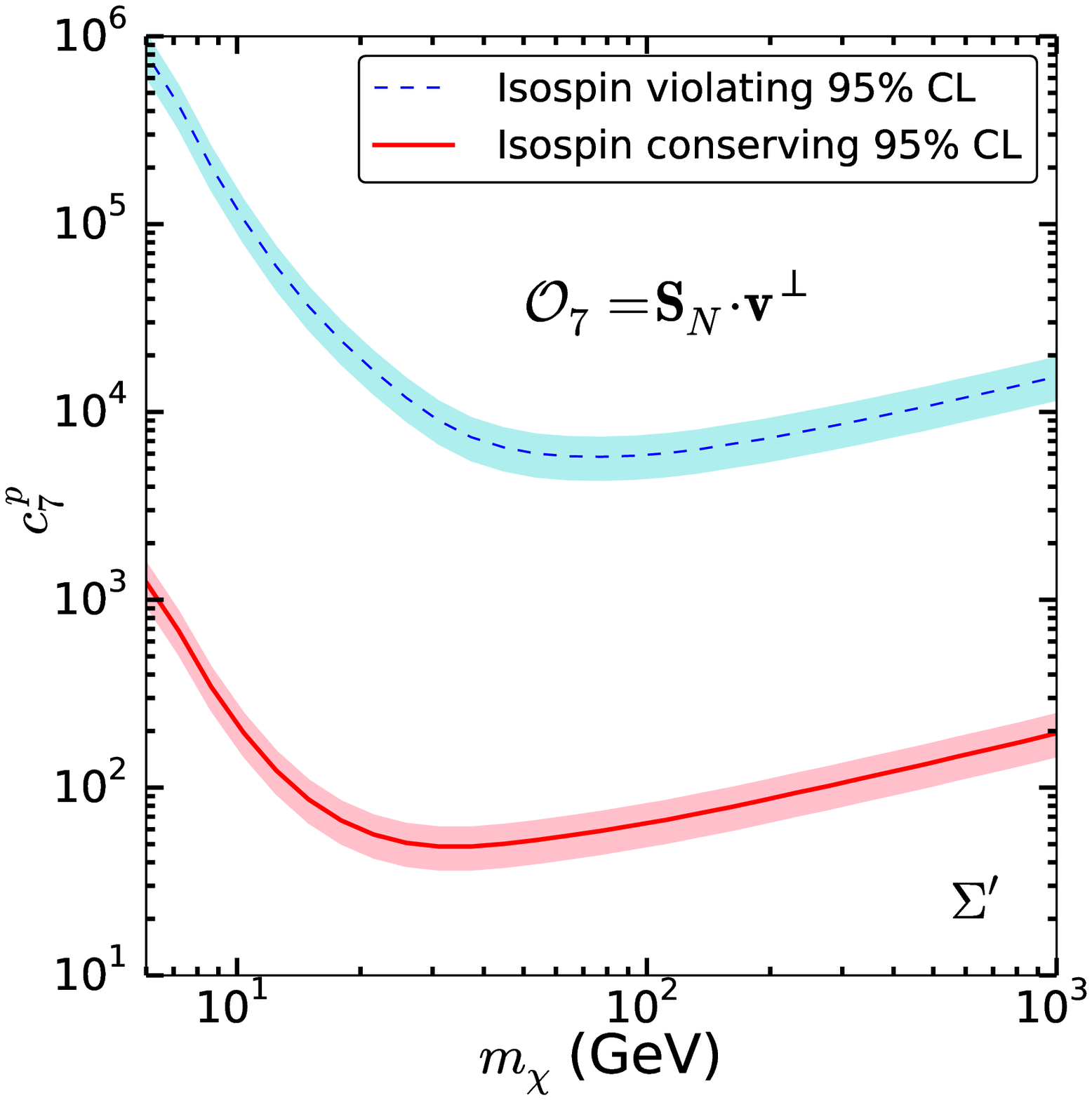}
\includegraphics[width=0.4\textwidth]{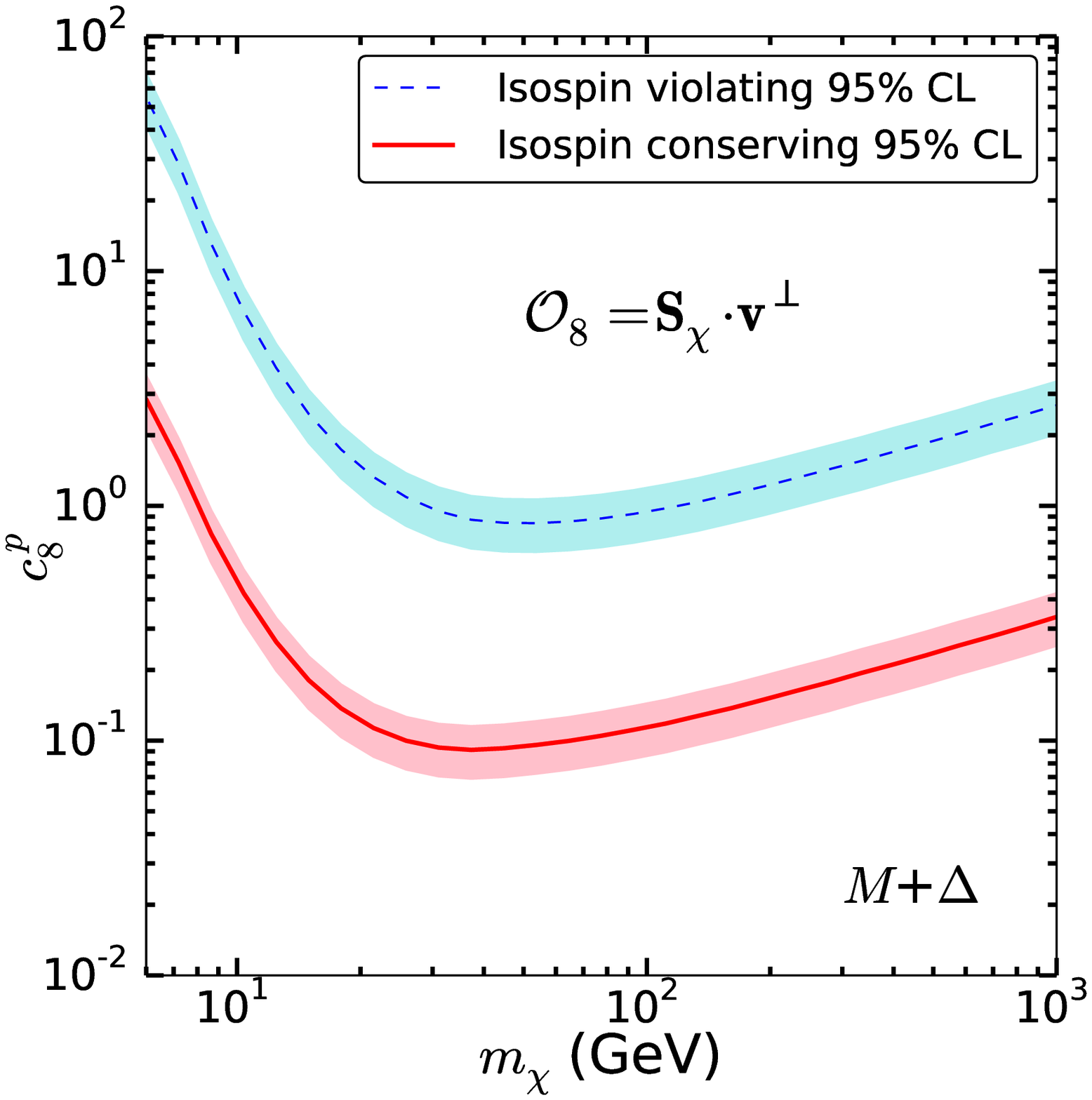}\\
\includegraphics[width=0.4\textwidth]{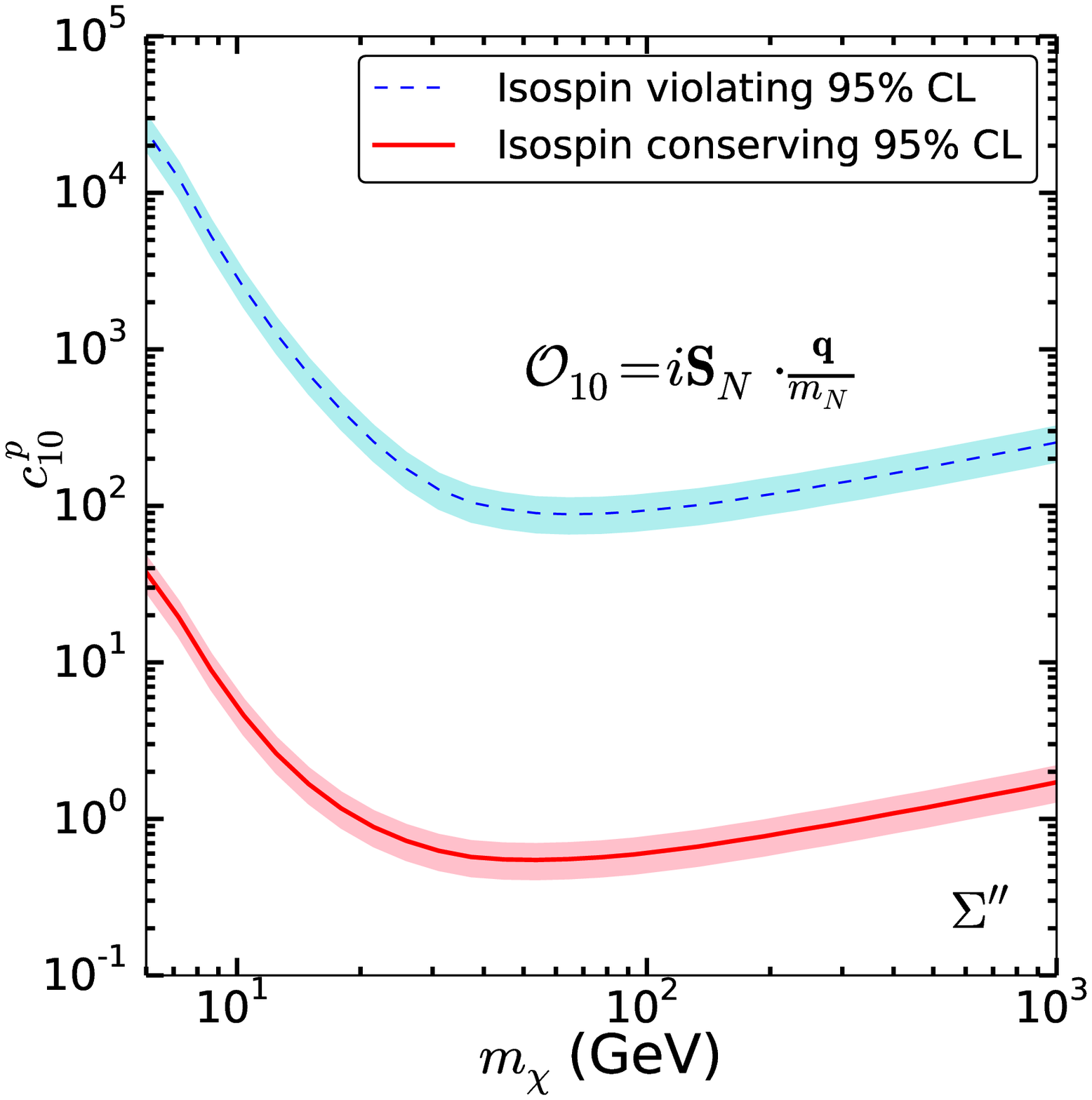}
\includegraphics[width=0.4\textwidth]{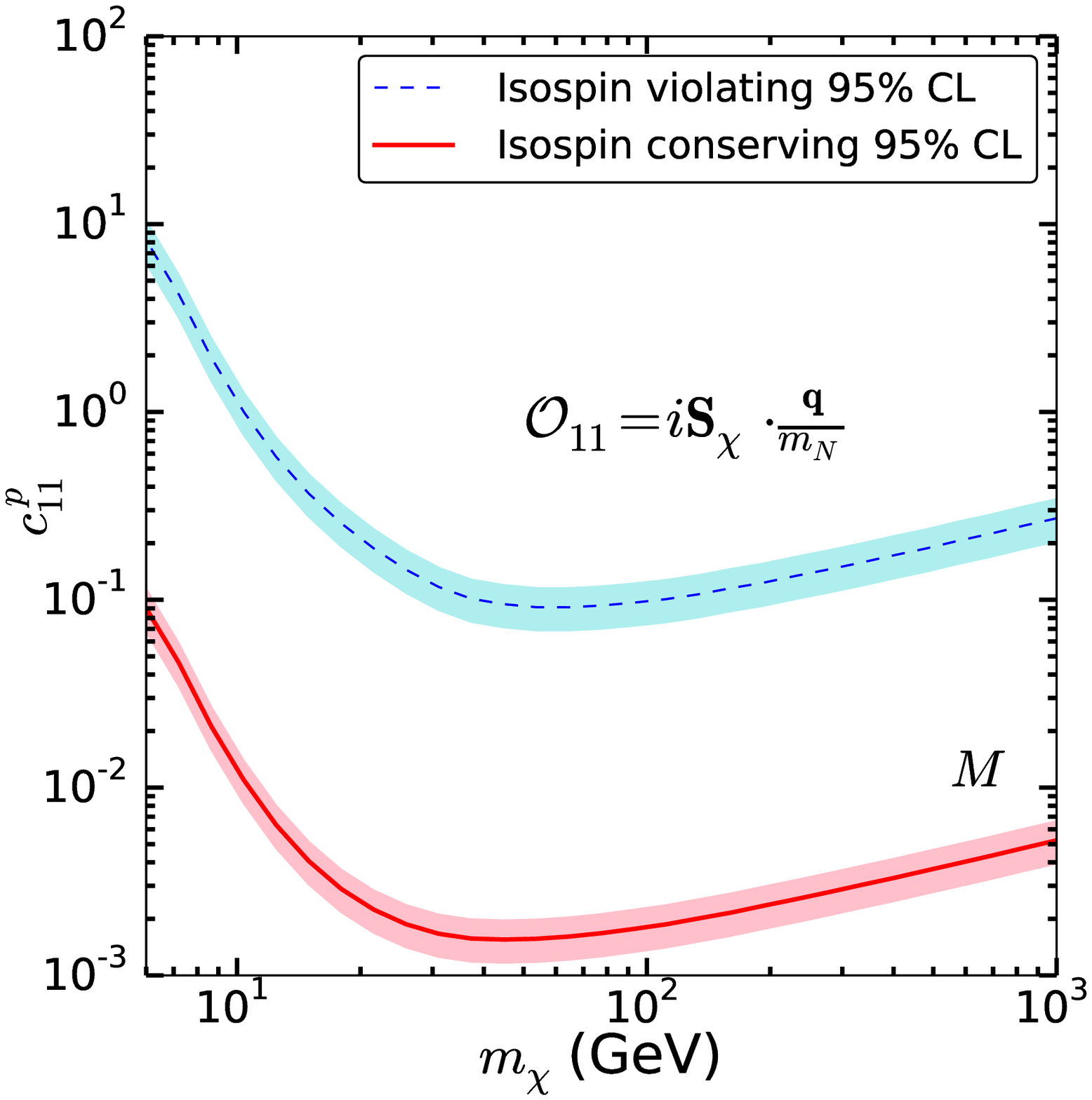}\\
\caption{The combined 95\% CL upper limits for non-relativistic operator
coefficients of $\mathcal{O}_1$, $\mathcal{O}_4$, $\mathcal{O}_7$, 
$\mathcal{O}_8$, $\mathcal{O}_{10}$, and $\mathcal{O}_{11}$. 
%In each panel, the PandaX (2016), LUX (2013-2016), and XENON1T data are 
%included. 
The red solid lines and blue dashed lines are for the ISC and ISV scenarios, 
respectively. The color bands demonstrate the uncertainties of the local
DM density measurements~\cite{Read:2014qva}. 
\label{fig:2opC}}
\end{figure}

In Fig.~\ref{fig:2opC}, the $95\%$ upper limits for the operators that 
consist of only two basis three-vectors given in Eq.~\eqref{Eq:fourvectors} 
are presented, including $\mathcal{O}_1$, $\mathcal{O}_4$, $\mathcal{O}_7$, 
$\mathcal{O}_8$, $\mathcal{O}_{10}$, and $\mathcal{O}_{11}$. 
The red solid lines are for the ISC scenario and blue dashed lines are 
for the ISV scenario. For convenience, we also label the response type by 
using the notations $M$, $\Delta$, $\Sigma^\prime$, $\Sigma^{\prime\prime}$, 
$\tilde{\Phi}^\prime$, and $\Phi^{\prime\prime}$, which refer to the DM 
current by vector charge, vector transverse magnetic, axial transverse 
electric, axial longitudinal, vector transverse electric, and vector 
longitudinal operators, respectively \cite{Anand:2013yka}.

Among the operators shown on Fig.~\ref{fig:2opC}, 
only $\mathcal{O}_1$ and $\mathcal{O}_4$ are DM velocity independent 
operators, which correspond to the traditional spin-independent and 
spin-dependent interactions, respectively. The constraints on the 
spin-independent cross section ($\mathcal{O}_1$) are stronger than the 
spin-dependent cross section ($\mathcal{O}_4$), well known from previous 
xenon-type experiments.
%in the allowed coupling size. 
%On the other hand, those 
The limits for the nucleon spin dependent operators ($\mathcal{O}_7$ and 
$\mathcal{O}_{10}$) are always weaker than the DM spin dependent operators 
($\mathcal{O}_8$ and $\mathcal{O}_{11}$). It can be seen from 
Eq.~\eqref{eq:amplitudes} and Eq.~\eqref{eq:Response}  that the operators  
$\mathcal{O}_1$, $\mathcal{O}_8$ and $\mathcal{O}_{11}$ are vector charge 
current interactions, and
%and have similar form of the DM and nuclear response functions. However, 
the operators $\mathcal{O}_4$, $\mathcal{O}_{7}$, and  
$\mathcal{O}_{10}$ are the axial current interactions.

\begin{figure}[!htp]
\includegraphics[width=0.45\textwidth]{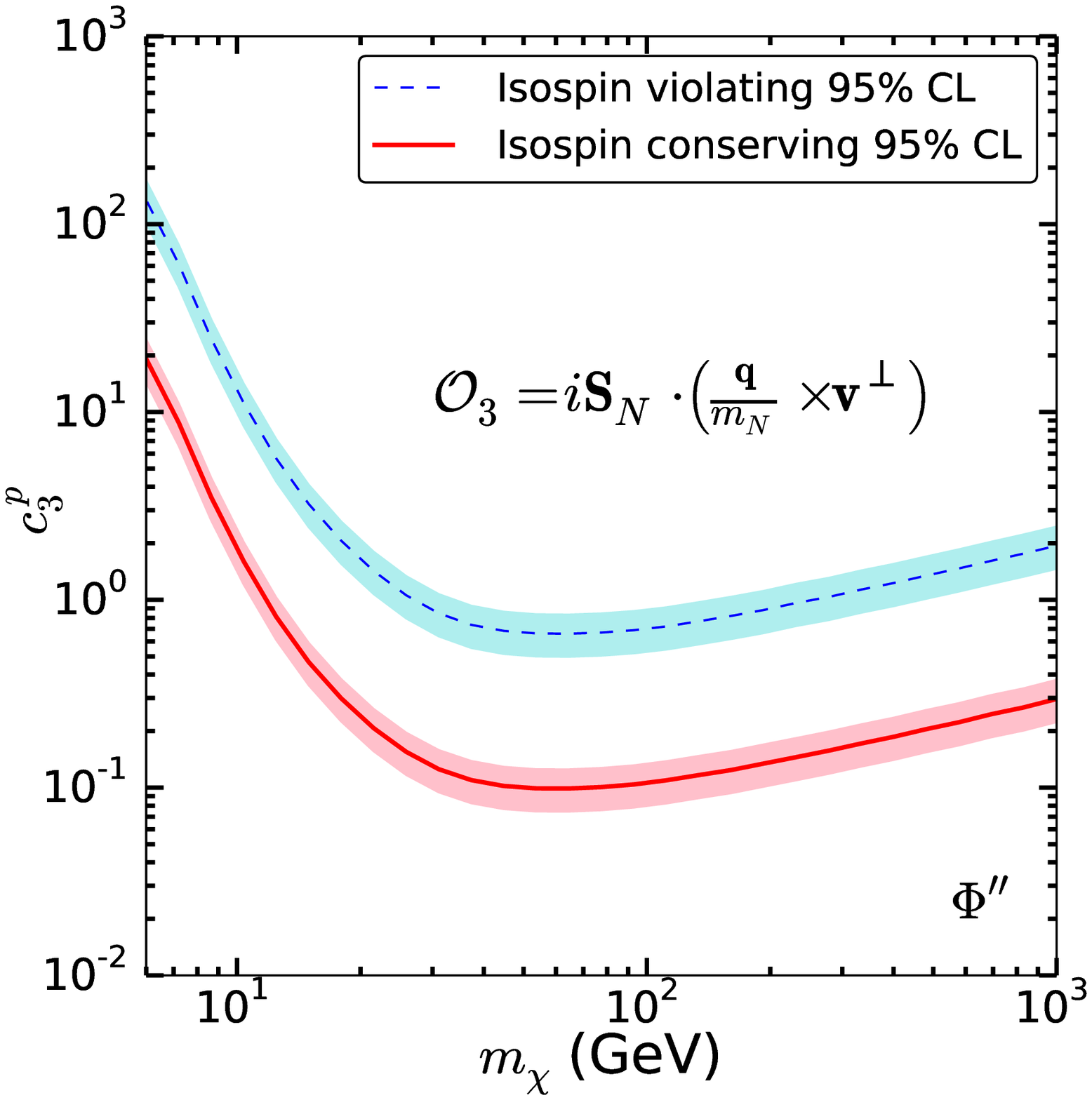}
\includegraphics[width=0.45\textwidth]{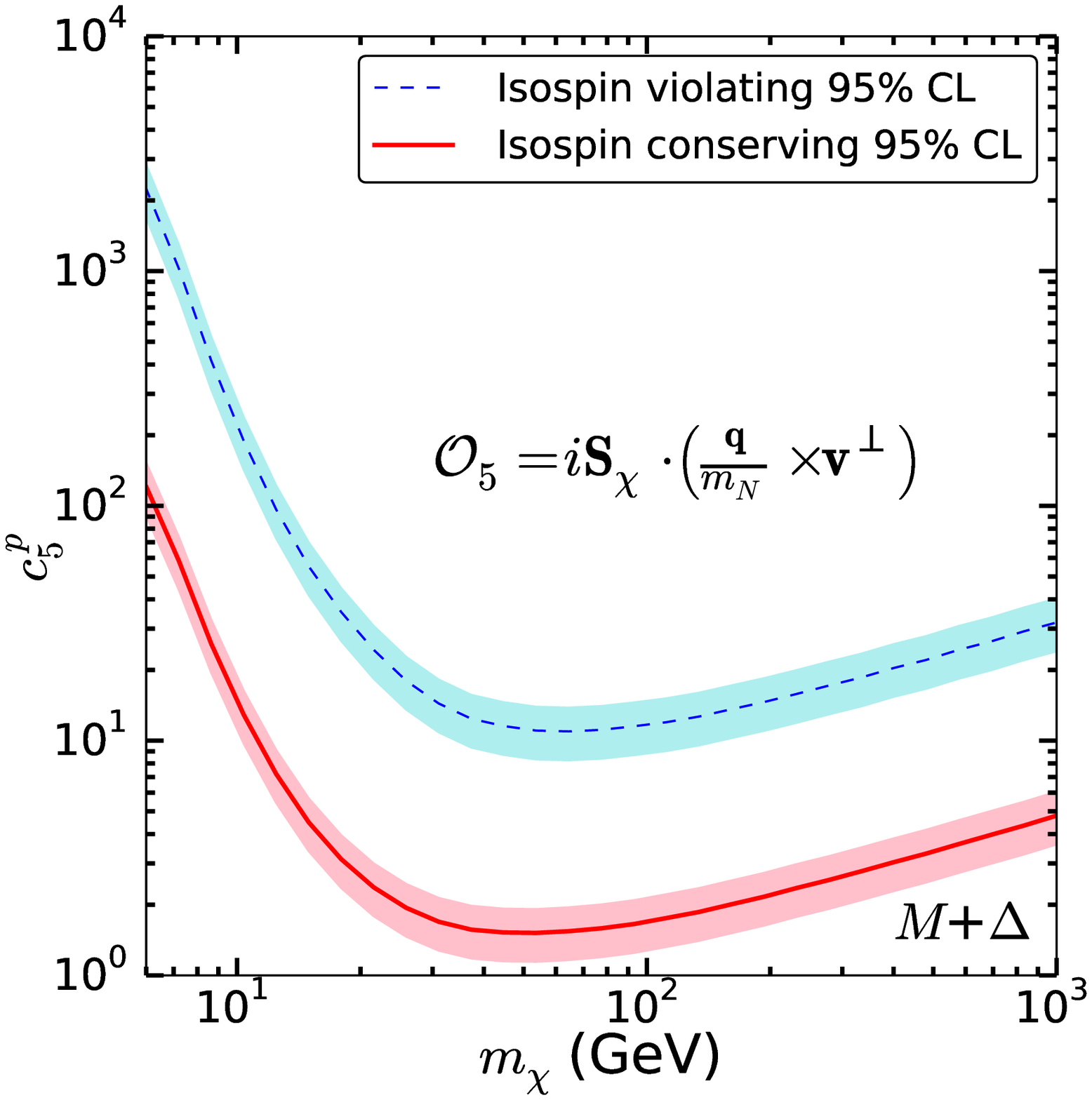}\\
\includegraphics[width=0.45\textwidth]{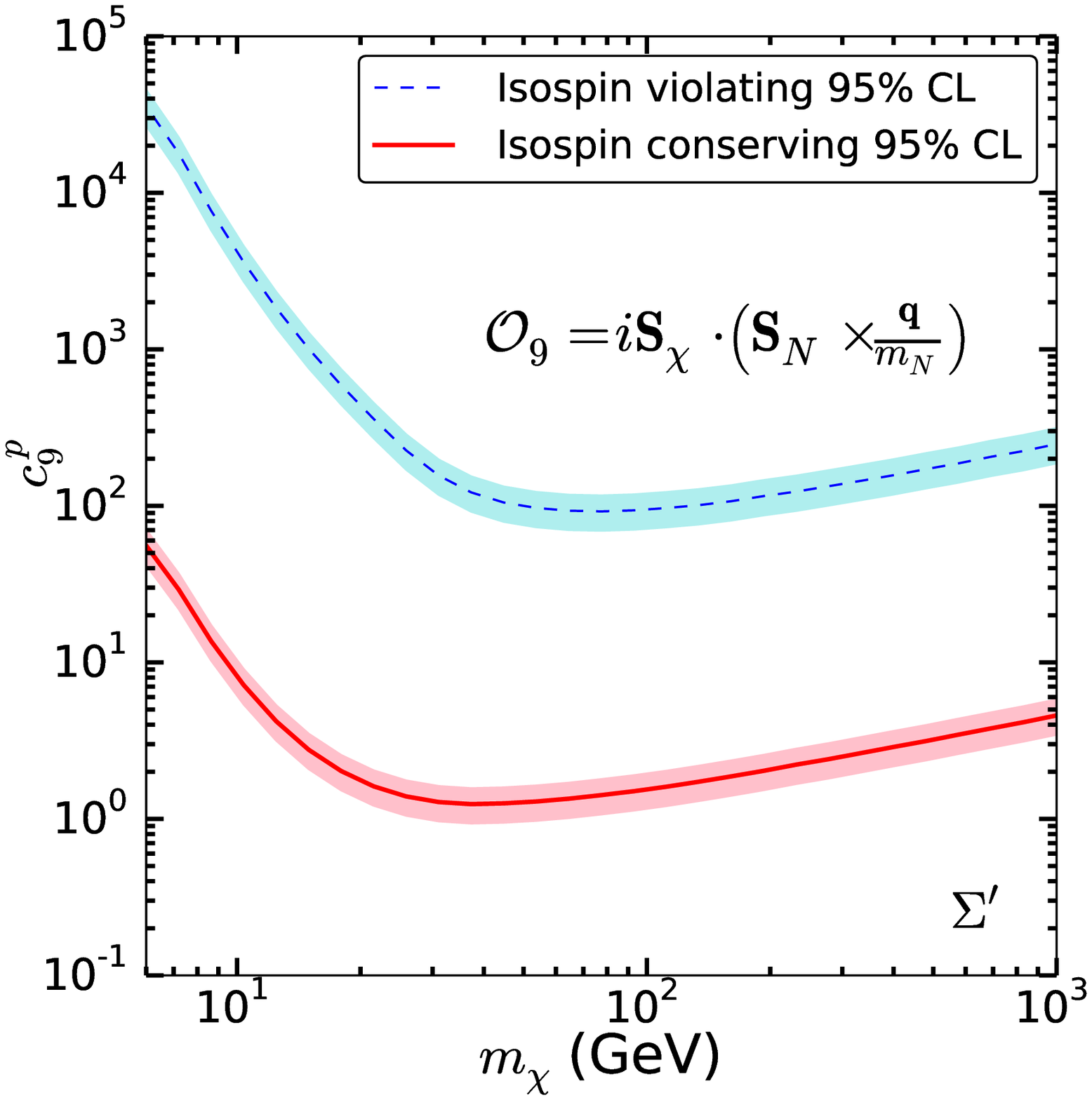}
\includegraphics[width=0.45\textwidth]{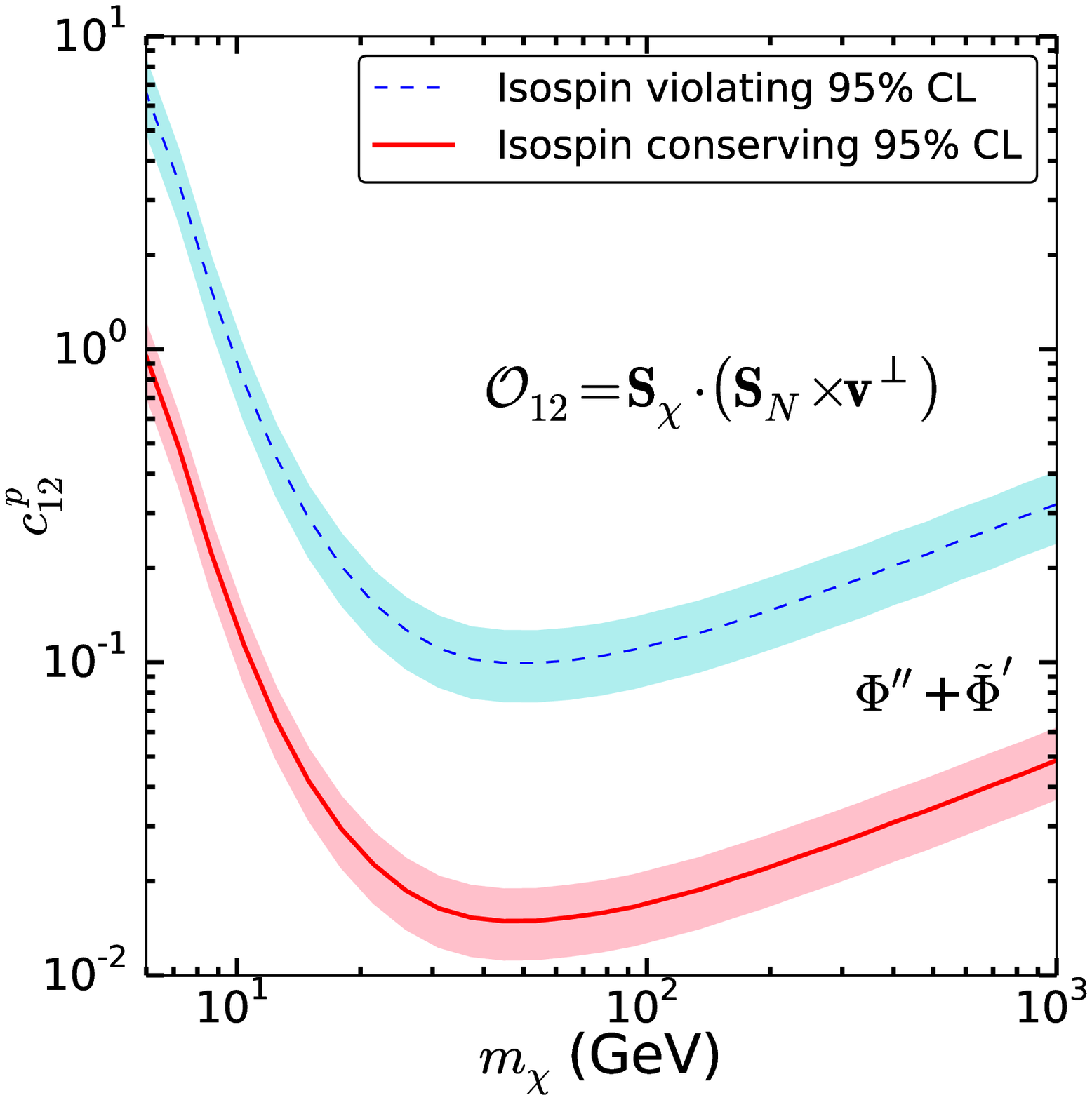}
\caption{Same as Fig.~\ref{fig:2opC} but for operators $\mathcal{{O}}_3$, 
$\mathcal{{O}}_5$, $\mathcal{{O}}_9$ and $\mathcal{{O}}_{12}$. 
\label{fig:3opC}}
\end{figure}

Fig.~\ref{fig:3opC} shows the constraints on operators that consist of 
three 3-vectors, including $\mathcal{{O}}_3$, $\mathcal{{O}}_5$, 
$\mathcal{{O}}_9$ and $\mathcal{{O}}_{12}$. In contrast to 
Fig.~\ref{fig:2opC}, the limits for nucleon spin dependent operator
($\mathcal{O}_3$, upper left panel of Fig.~\ref{fig:3opC}) are stronger 
than the DM spin dependent one ($\mathcal{O}_{5}$, upper right panel).
% of Fig.~\ref{fig:3opC} whose cross 
%sections mainly formed by the products of DM/nuclear spin and 
%${\bf{{q}}}\times{\bf{{v}}}^{\perp}$, the limit for nuclear spin 
%dependent operator ($\mathcal{O}_3$) is stronger than DM spin dependent 
%operator ($\mathcal{O}_5$).
%\mkred{(Zuowei: why the limit on $\mathcal{O}_3$ is stronger than $\mathcal{O}_5$?)}
This is because that the contribution of the nuclear response function 
$\Phi^{\prime\prime}$ is larger than $M$, since $\Phi^{\prime\prime}$ 
possess not only the scalar contribution but also the quasicoherent 
one~\cite{Fitzpatrick:2012ix,Anand:2013yka}.  For a similar reason, the 
limits for $\mathcal{O}_{12}$ is two orders of magnitude stronger than 
that of $\mathcal{O}_9$, since the operator $\mathcal{O}_{12}$ 
also has the quasicoherent contribution from $\Phi^{\prime\prime}$.

\begin{figure}[!htb]
\includegraphics[width=0.45\textwidth]{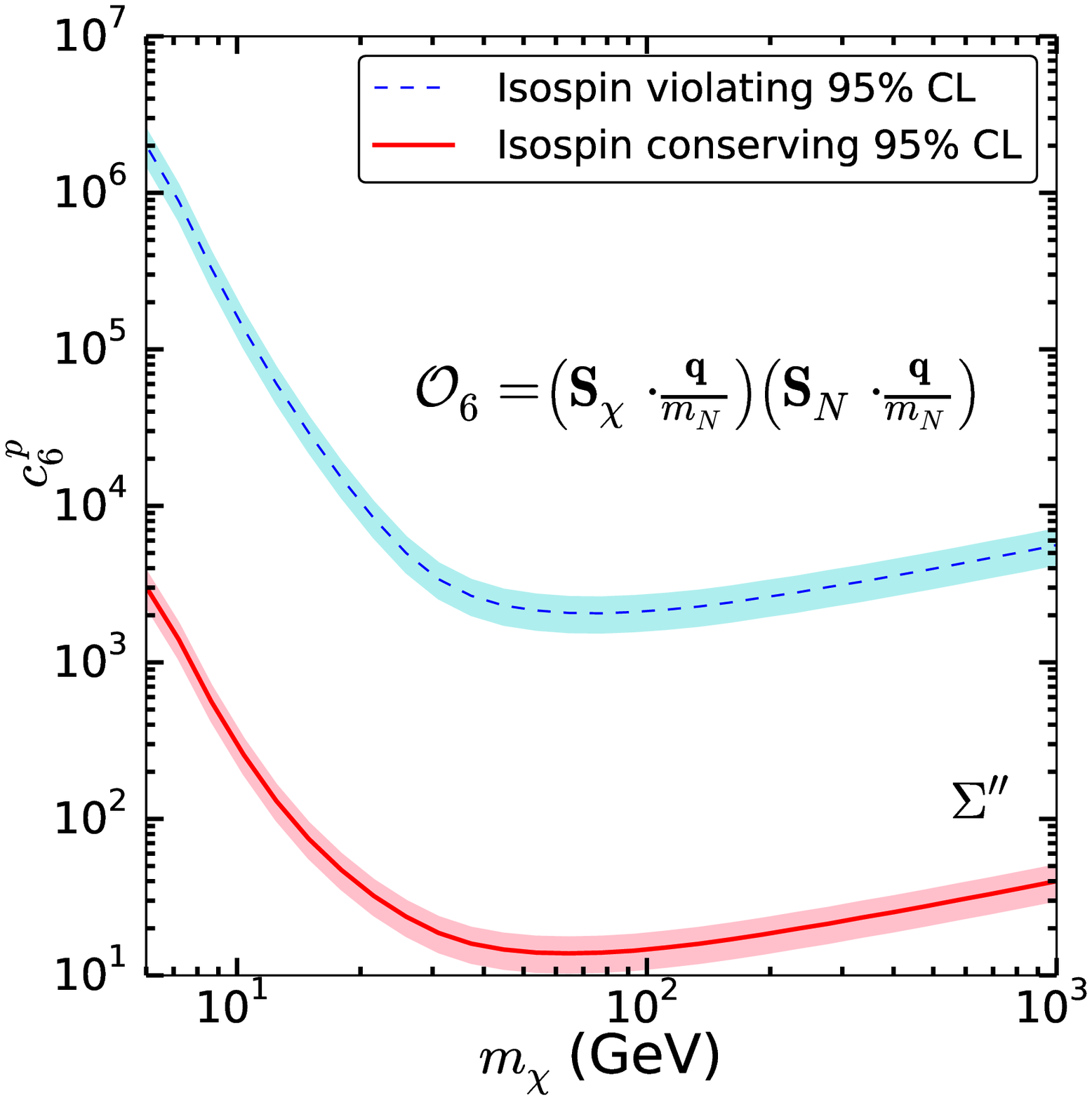}
\includegraphics[width=0.45\textwidth]{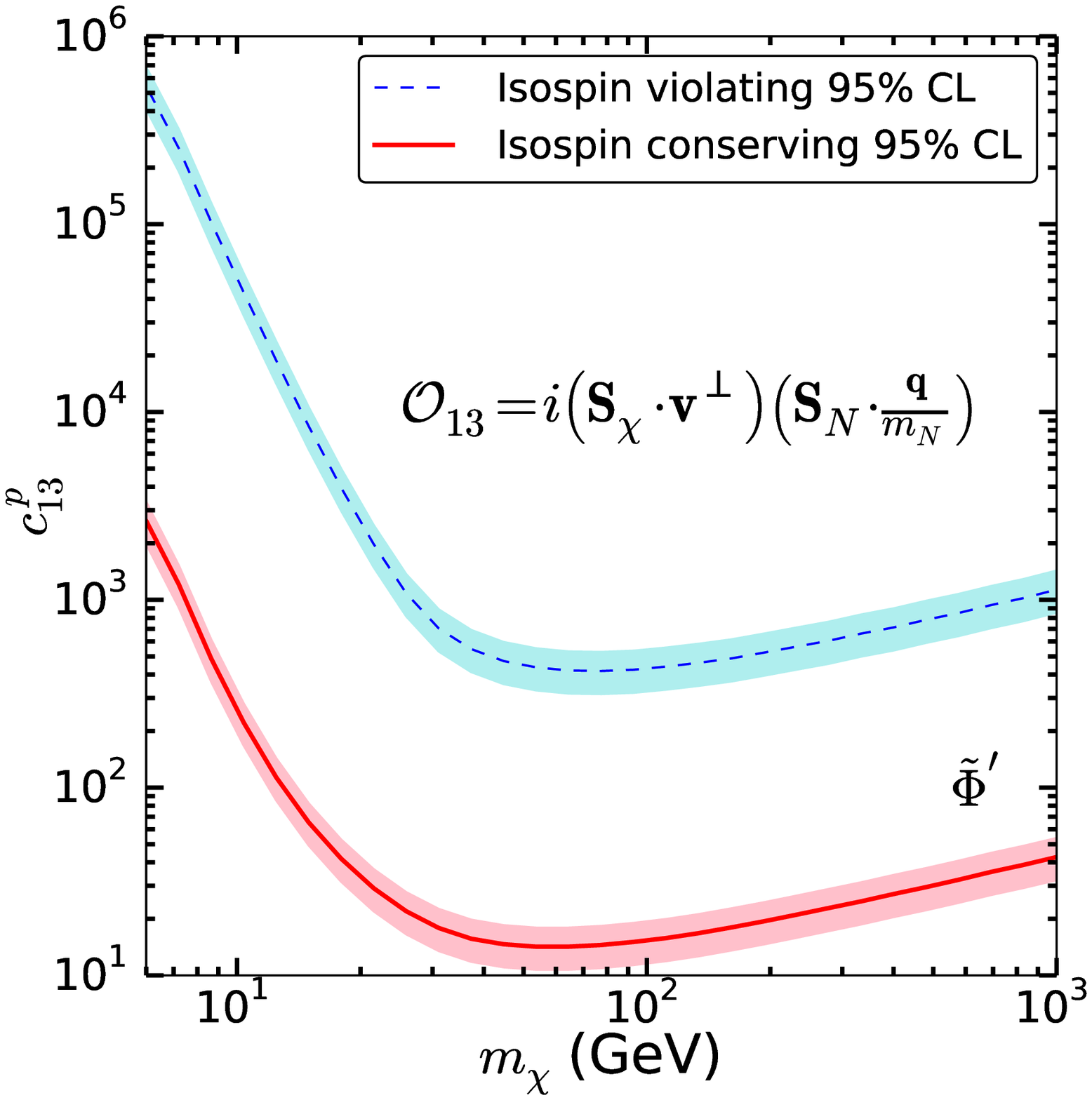}\\
\includegraphics[width=0.45\textwidth]{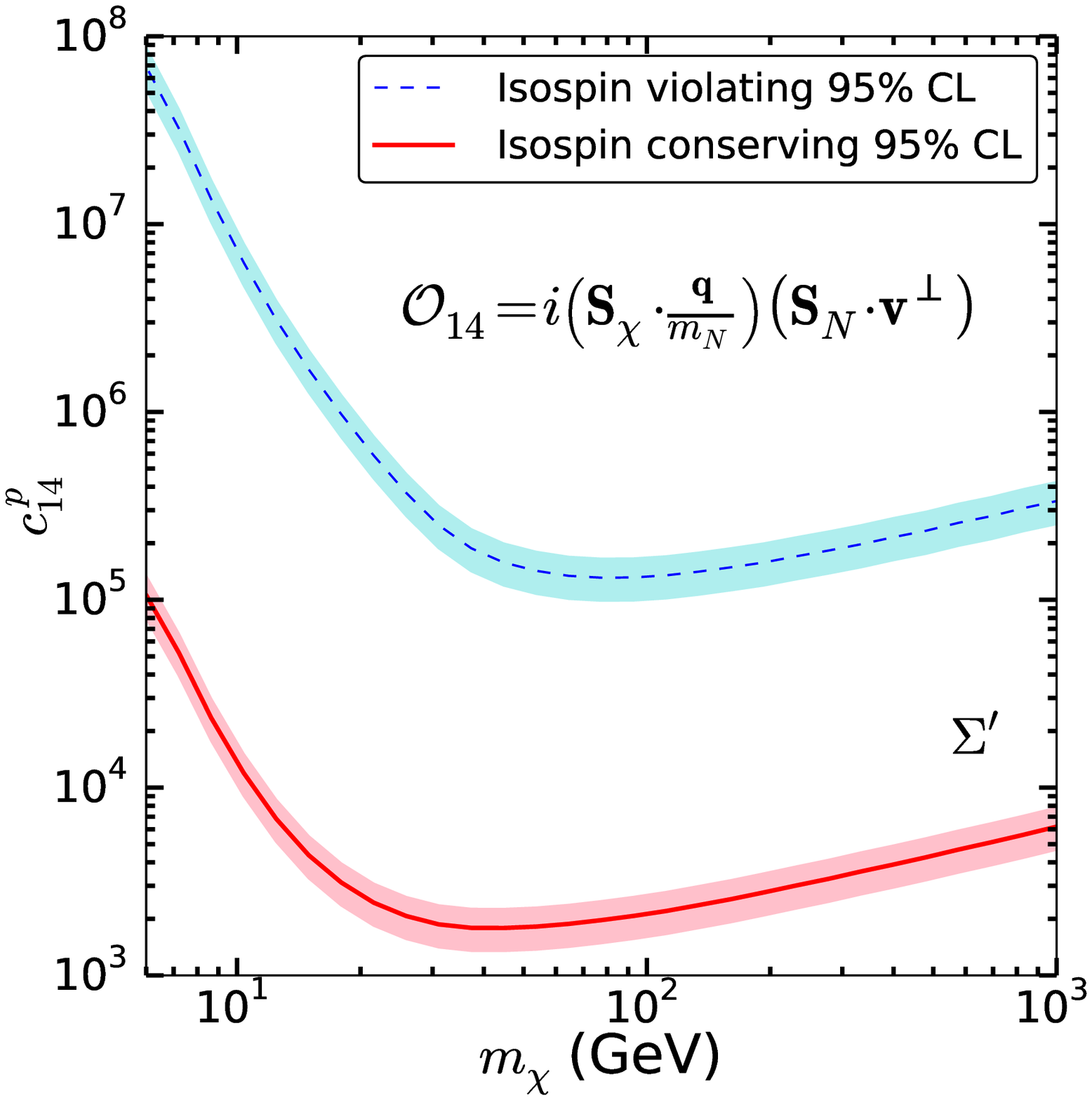}
\includegraphics[width=0.45\textwidth]{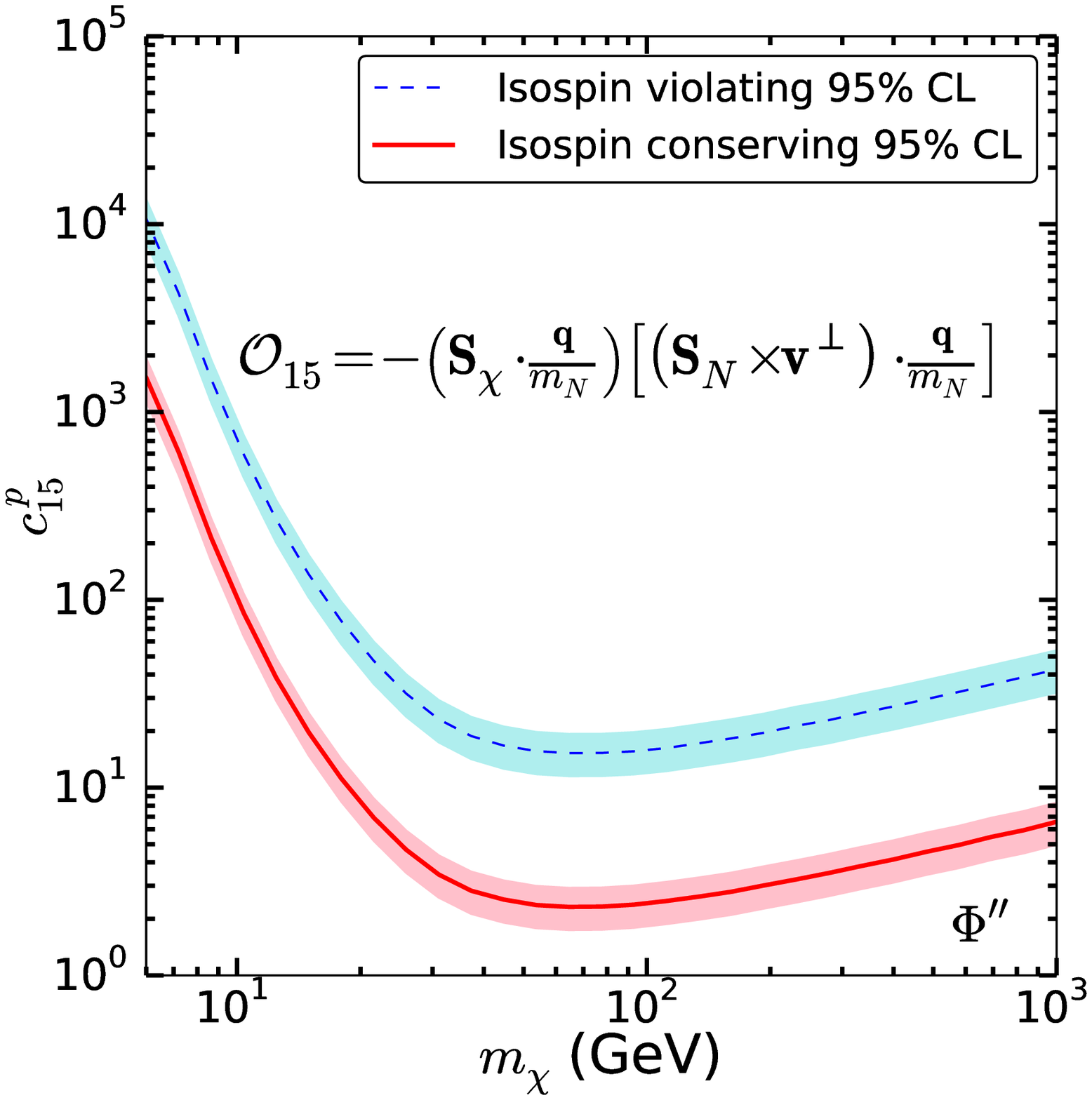}
\caption{Same as Fig.~\ref{fig:2opC} but for operators $\mathcal{{O}}_{6}$, 
$\mathcal{{O}}_{13}$, $\mathcal{{O}}_{14}$ and $\mathcal{{O}}_{15}$. 
%In each panel, the PandaX (2016), LUX (2014-2016), and XENON1T data are 
%included. The red lines and blue dashed lines are isospin conserving and 
%violating, respectively. The color bands demonstrate the astrophysical 
%uncertainties from the different approaches~\cite{Read:2014qva}.  
\label{fig:4opC}}
\end{figure}

The results for the rest four operators, $\mathcal{{O}}_{6}$, 
$\mathcal{{O}}_{13}$, $\mathcal{{O}}_{14}$, and $\mathcal{O}_{15}$,
are shown in Fig.~\ref{fig:4opC}. Except $\mathcal{O}_{15}$ which 
includes 5 basis vectors, the operators $\mathcal{{O}}_{6}$, 
$\mathcal{{O}}_{13}$, and $\mathcal{{O}}_{14}$ have 4 basis vectors
combined. 
The ISC limits are similar for $\mathcal{{O}}_{6}$ and $\mathcal{{O}}_{13}$,
 while the ISV limits differ by about 
one order of magnitude for these two operators. Even though $\mathcal{{O}}_{14}$ has similar 
structure as $\mathcal{{O}}_{6}$ and $\mathcal{{O}}_{13}$, it behaves 
differently. As pointed out in Ref.~\citep{Anand:2013yka}, 
the structure ${\bf{S}}_{N}\cdot {\bf{v}}^{\perp}$, 
where the DM couples to nucleon via the axial charge, 
can have a vanishing intrinsic velocity contribution so that 
it leads to standard spin-dependent operators.
In addition, for the 5-vector combination $\mathcal{O}_{15}$, 
we found that the difference between ISC and ISV are the smallest 
among the 14 operators. 
This is also expected because the interference terms 
in $\mathcal{O}_{15}$ are from high dimensional operators.

\begin{figure}[!htb]
\includegraphics[width=0.55\textwidth]{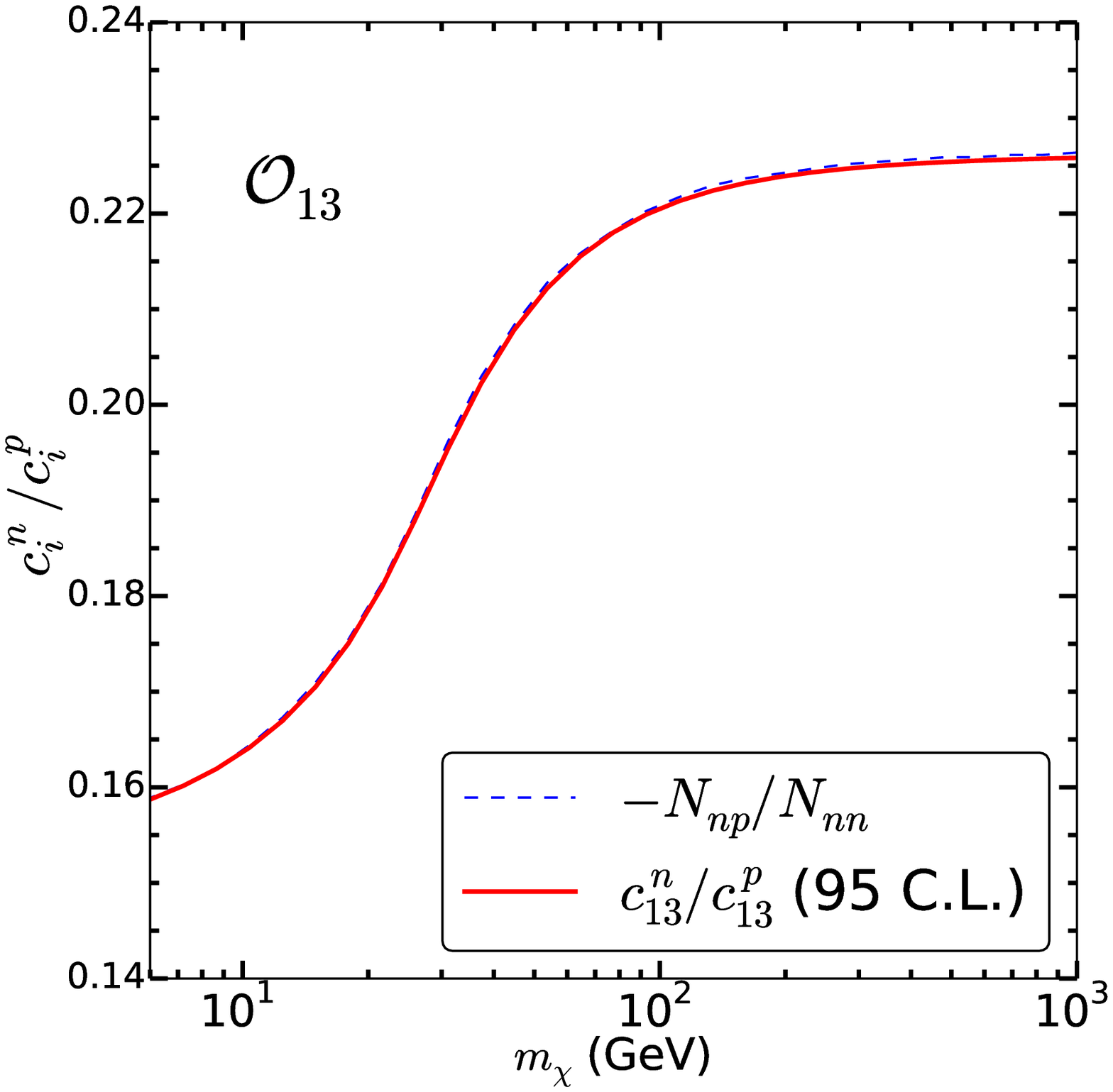}
\caption{The predicted ratio of the coefficient $c_{13}^n$ to $c_{13}^p$ 
by the minimum condition (blue dashed line) and the maximum cancellation 
at the $95\%$ CL (red line). 
\label{fig:ISV13}}
\end{figure}

Finally, we want to discuss the ISV coupling ratio for maximum cancellation. 
In order to study the interference between the $\chi$-$n$ and $\chi$-$p$ 
amplitudes, we introduce an interference parameter $N_{np}$ and rewrite 
the total event number in a matrix form, as
\begin{eqnarray}
N(c_i^n, c_i^p) =
\left(\begin{array}{cc}
     c_i^n & c_i^p 
\end{array}\right)
\left[\begin{array}{cc}
     N_{nn}              &  N_{np}                  \\
     N_{np}             &   N_{pp}                  \\
\end{array}\right]
\left(\begin{array}{ccc}
     c_i^n     \\
     c_i^p    \\
\end{array}\right) \, , 
\end{eqnarray}  
where $N_{nn}$ is the predicted event number with $c_i^n=1$ and $c_i^p=0$,
and $N_{pp}$ is the one with $c_i^n=0$ and $c_i^p=1$. The interference 
$N_{np}$ can be easily obtained from the numerical computation. 
Applying the minimum condition for the variable $c_i^n/c_i^p$, the minimum 
value of $N$ (maximum cancellation) is located at 
\begin{equation}\label{eq:maxISV}
\frac{c_i^n}{c_i^p}=-\frac{N_{np}}{N_{nn}}.
\end{equation}
Such a coupling ratio is determined at the event level, which can slightly 
vary for different targets, cut efficiencies, and DM velocity distributions. 
In Fig.~\ref{fig:ISV13}, we present the predicted ratio of the coefficient 
$c_{13}^n$ to $c_{13}^p$ by the minimum condition (blue dashed line), 
compared with the maximum cancellation at the $95\%$ CL (red solid line) 
from Fig.~\ref{fig:4opC}. The difference between these two is very small, 
suggesting that the value $-N_{np}/N_{nn}$ gives an excellent estimate
of the ISV coupling ratio. One may note that the ratio becomes a constant 
of $0.225$ for $\mathcal{O}_{13}$ if the DM mass is much heavier than 
$100\gev$.

\begin{figure}[!htb]
\includegraphics[width=0.45\textwidth]{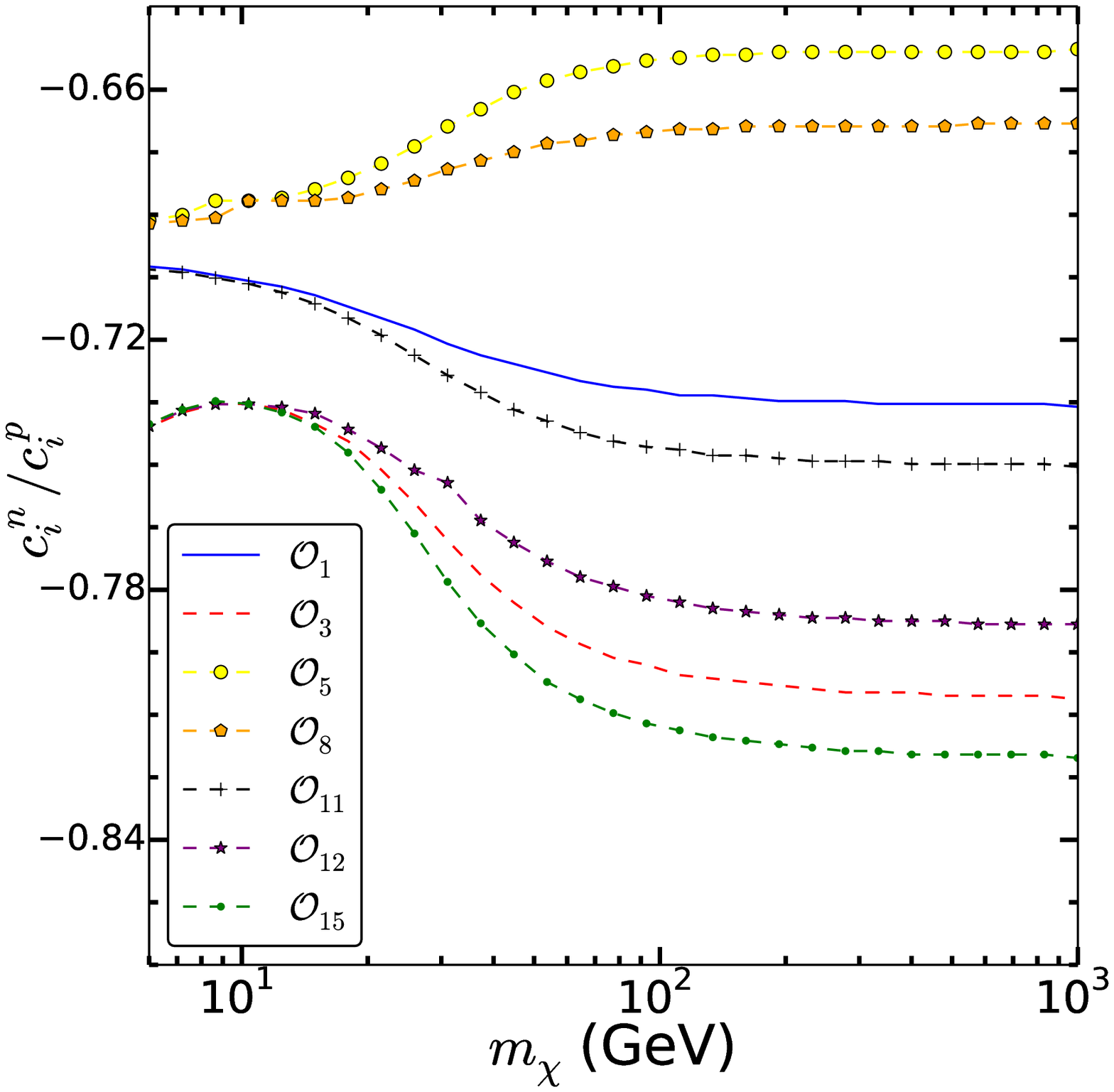}
\includegraphics[width=0.45\textwidth]{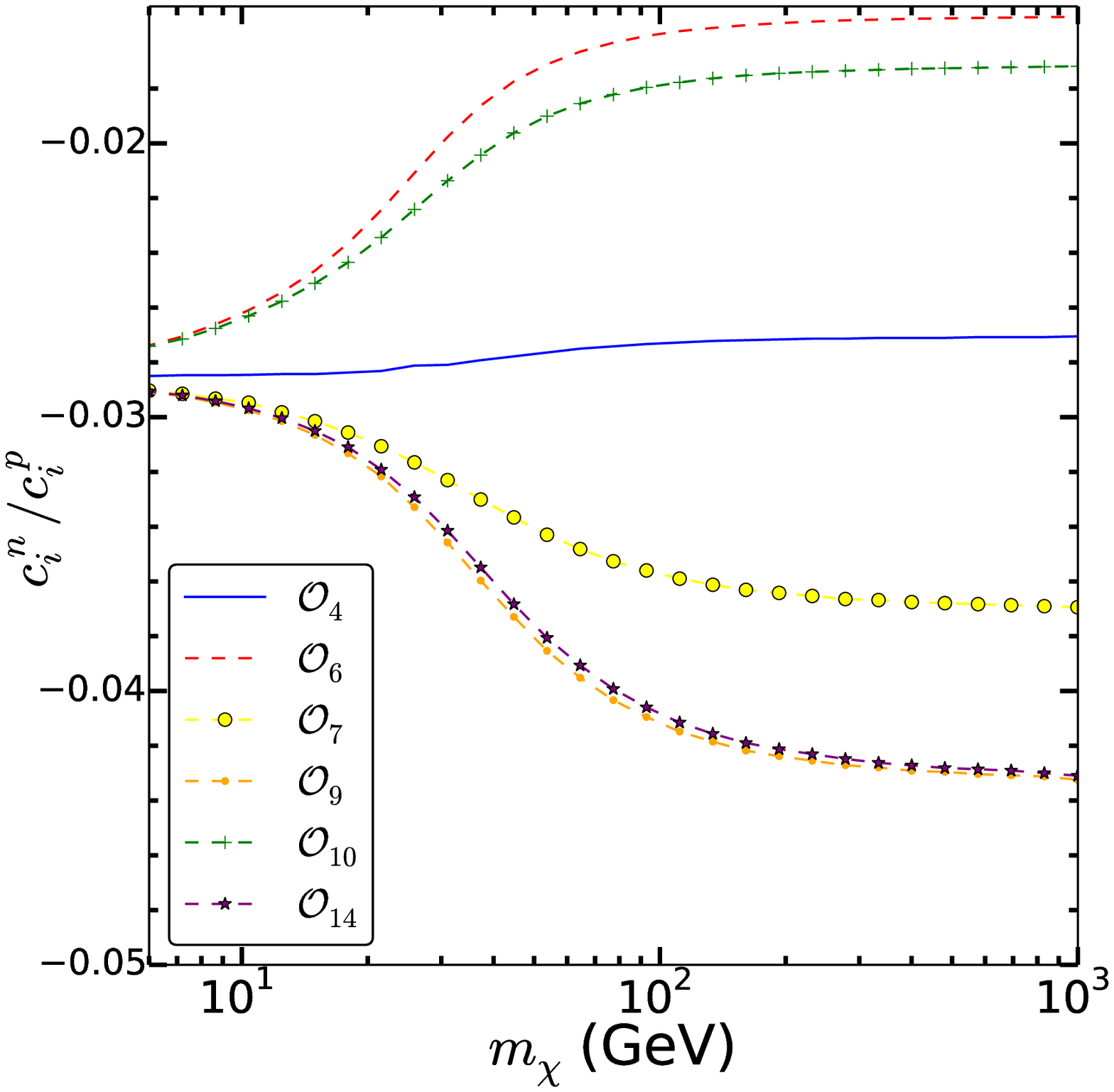}
\caption{The ratios the coefficients $c_i^n$ to $c_i^p$ for the other
13 operators except $\mathcal{O}_{13}$.   
\label{fig:ISV1and2}}
\end{figure}

Fig.~\ref{fig:ISV1and2} shows the ratios $c_i^n/c_i^p$ for the other 13
operators except $\mathcal{O}_{13}$. We can find that such ratios for
all the operators approach constant at high $\mchi$ regions.
%We found that the ratios of $c_i^n$ to $c_i^p$ for operators 
%$\mathcal{O}_1$ and $\mathcal{O}_4$ are relatively mildly changed 
%with respective to DM mass but all of operators achieve to a constant 
%at the larger $\mchi$. 
The ISV coupling ratio $c_i^n/c_i^p$ is between $-0.82$ to $-0.62$ for 
the operators shown in the left panel of Fig.~\ref{fig:ISV1and2}, and 
between $-0.043$ to $-0.015$ for the operators shown in the right panel. 
From Eq.~\eqref{eq:maxISV}, we can see that the sign of $c_i^n/c_i^p$ 
for the maximum cancellation is always determined by the interference 
term $N_{np}$. Interestingly, except for the operator $\mathcal{O}_{13}$ 
where the interference is negative, the interference of all the other operators 
is positive so that $c_i^n$ and $c_i^p$ have opposite signs 
for maximum ISV cancellations.

%\mkred{We emphasize that the two coupling coefficients and their likelihood 
%functions can still be affected by the velocity distribution at the low 
%$\mchi$ region. From Eq.~\eqref{eq:maxISV} we can find that the values of 
%$\mchi$, $v_0$, and $v_{esc}$ would still change the ratio a little bit.   
%One has to bear in mind that the ratios reported in Figs.~\ref{fig:ISV13}
%and \ref{fig:ISV1and2} are just required values for the maximum cancellation 
%to obtain the maximum $c_p$ values of the PandaX+LUX+XENON1T likelihood. 
%Any other ratio cannot have larger $c_p$, and these curves cannot be 
%treated as physical properties of the DM. (??QY: this paragraph is
%slightly difficult to be understood. Is is really necessary??)} 
%Our value can be provided to model builders for the reference.   

\subsection{Relativistic effective Lagrangians}

\begin{figure}[!htb]
\includegraphics[width=0.45\textwidth]{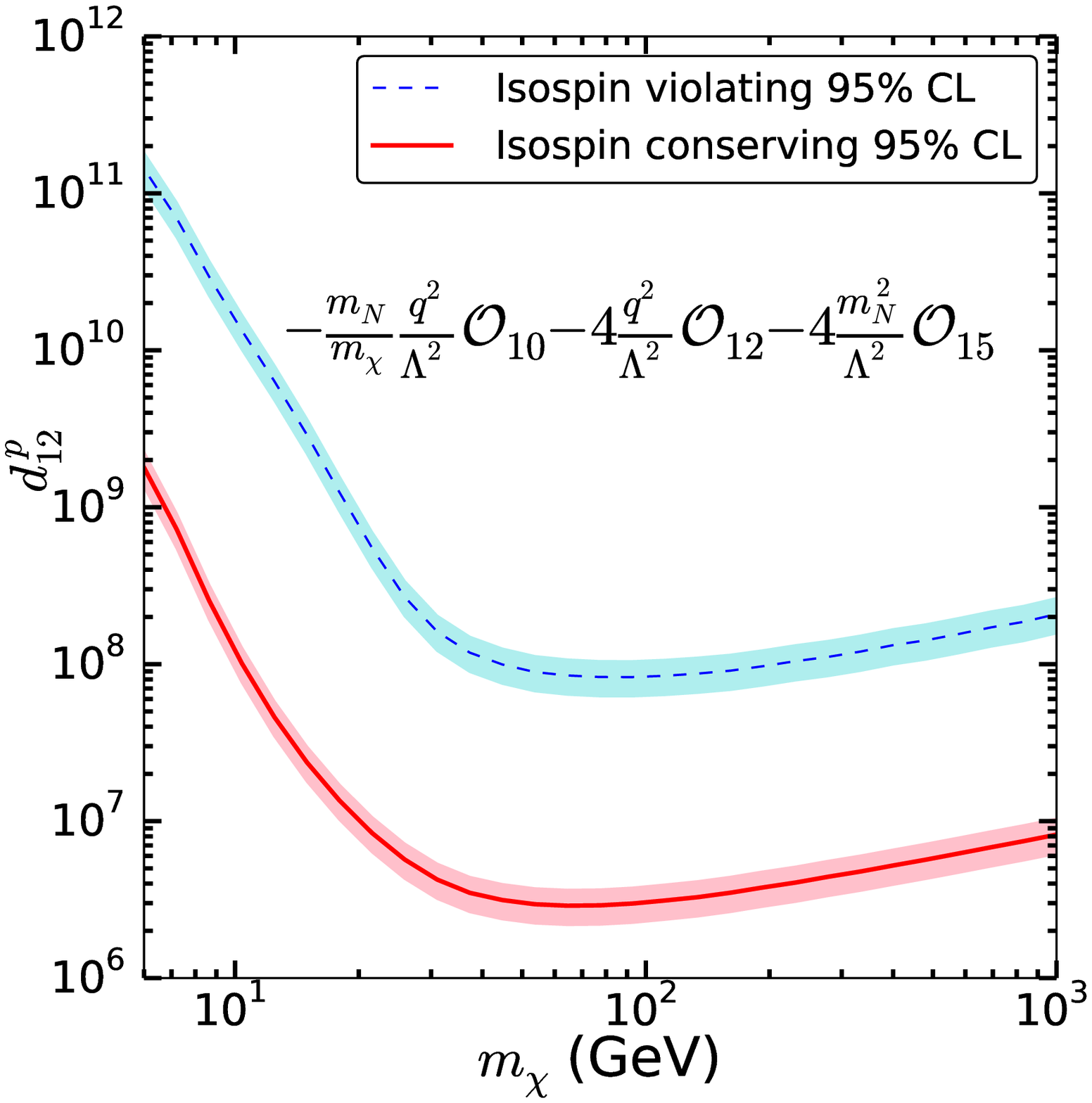}
\includegraphics[width=0.45\textwidth]{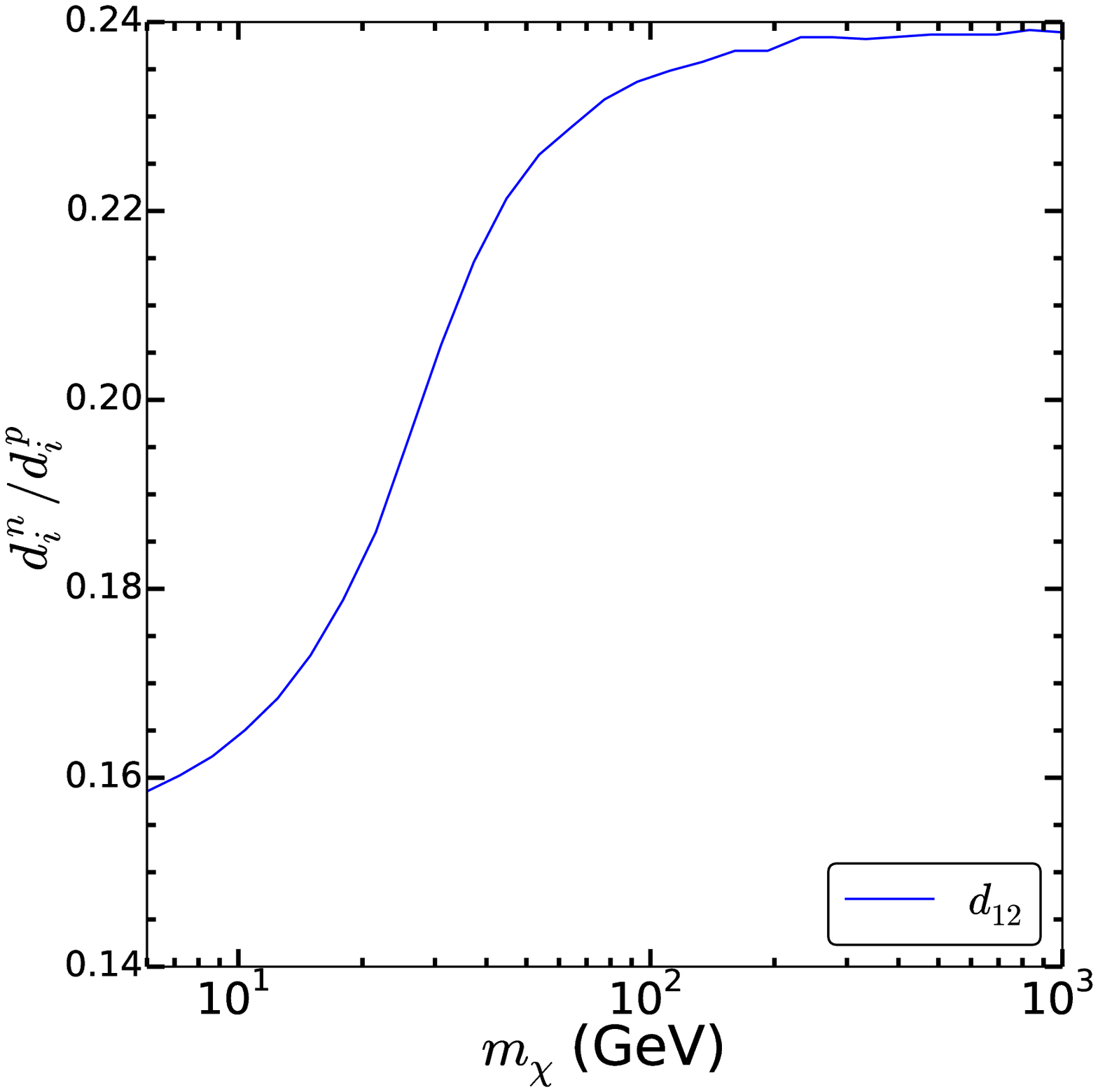}\\
\caption{Left panel: the 95\% CL upper limits of the ISC (red solid) 
and ISV (blue dashed) cases for the coupling $d^p_{12}$. 
The color bands demonstrate the uncertainties of the local DM density
measurements~\cite{Read:2014qva}. 
Right panel: the maximum cancellation ratio as a function of $\mchi$
for $\mathcal{L}_{12}$.   
\label{fig:d12}}
\end{figure}

\begin{figure}[!htb]
\includegraphics[width=0.45\textwidth]{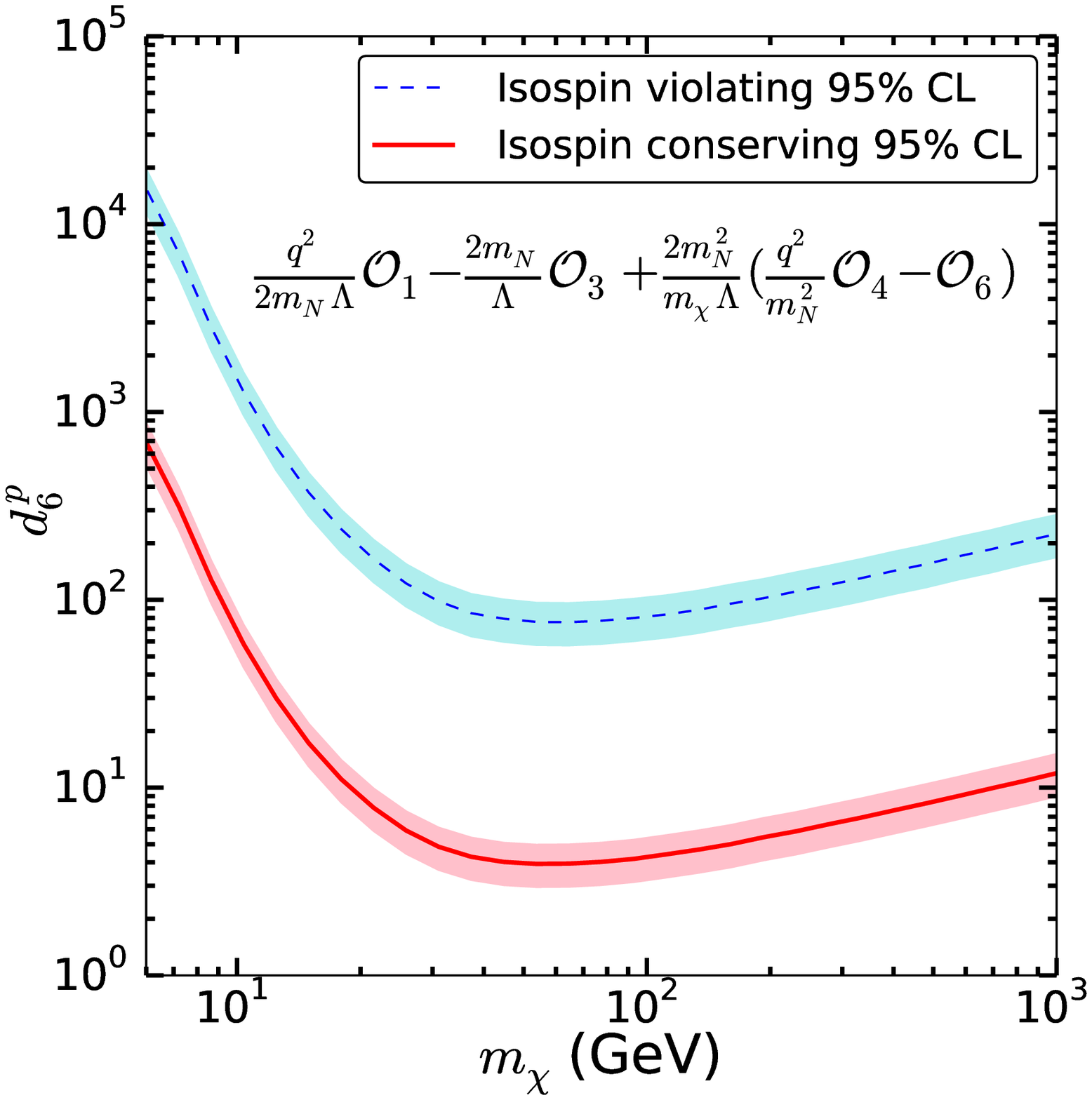}
\includegraphics[width=0.45\textwidth]{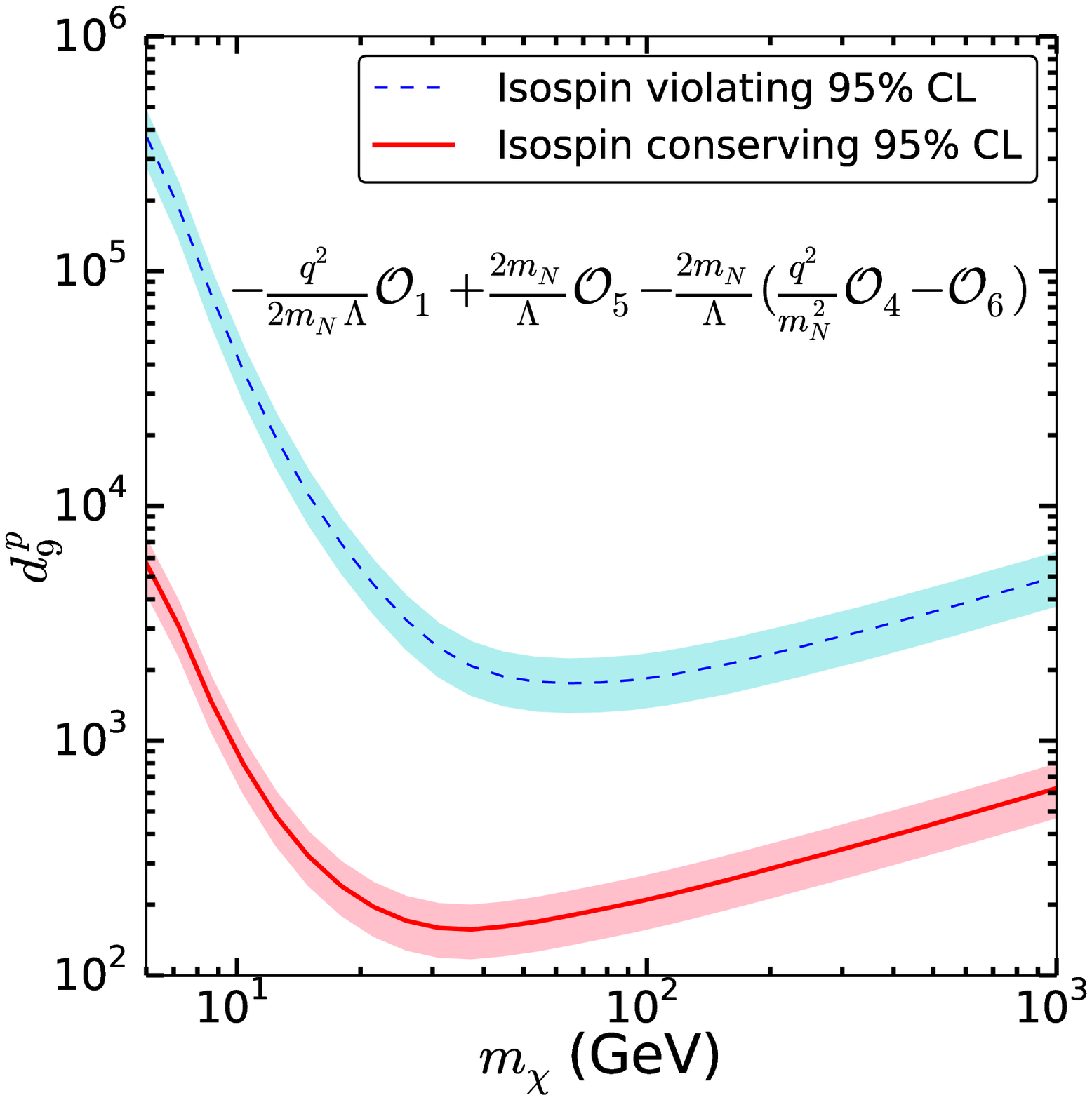}\\
\includegraphics[width=0.45\textwidth]{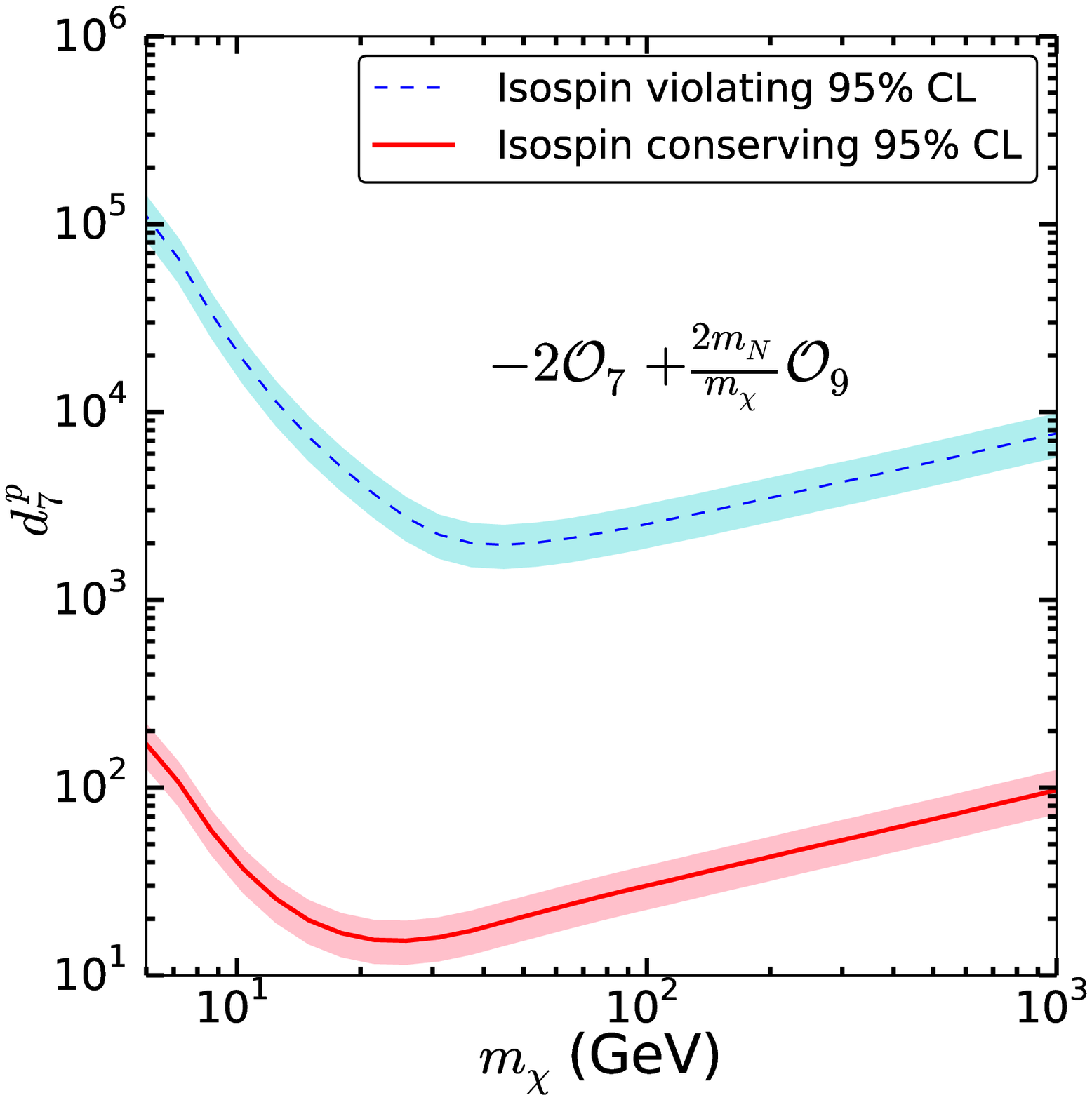}
\includegraphics[width=0.45\textwidth]{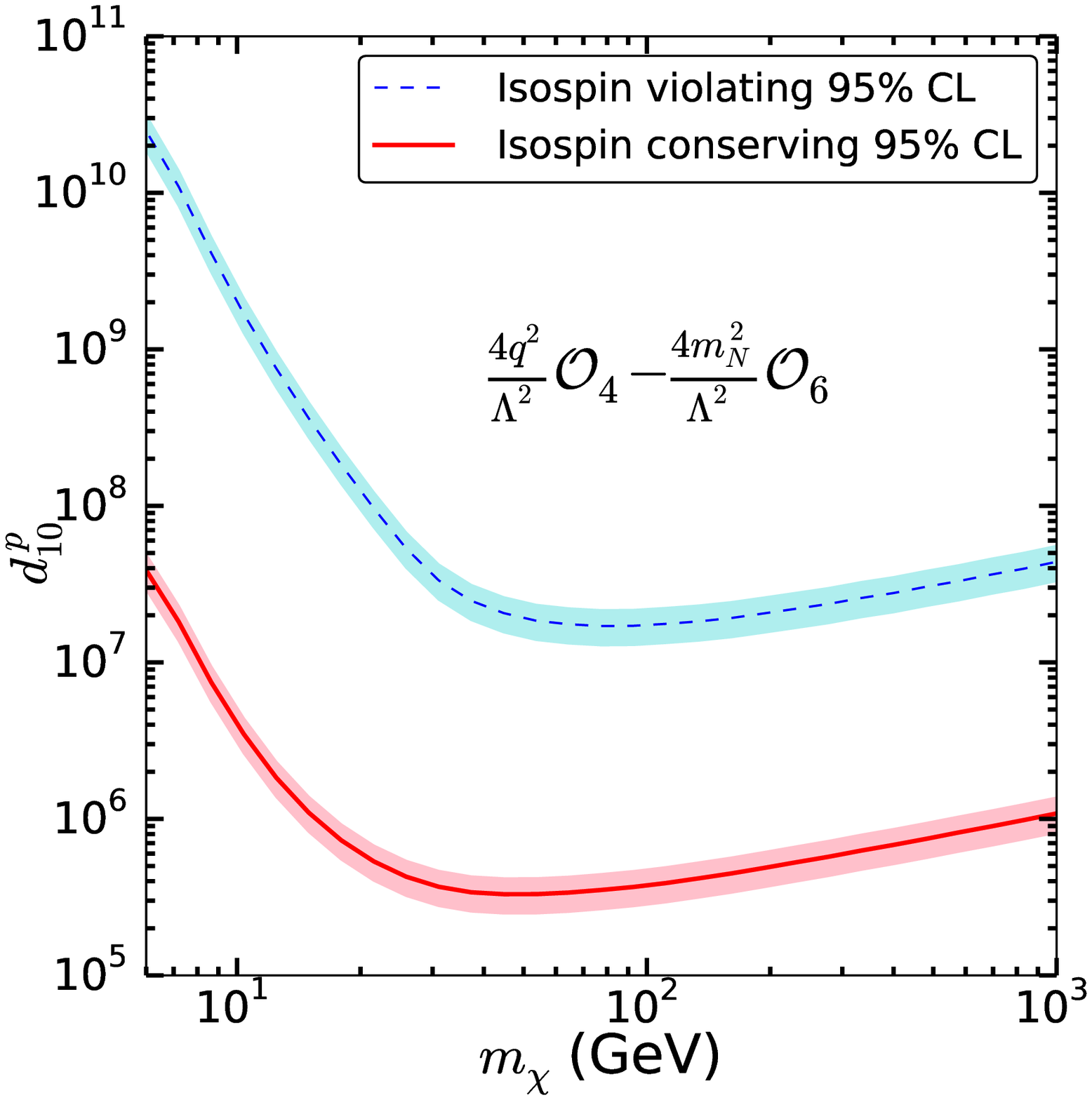}\\
\includegraphics[width=0.45\textwidth]{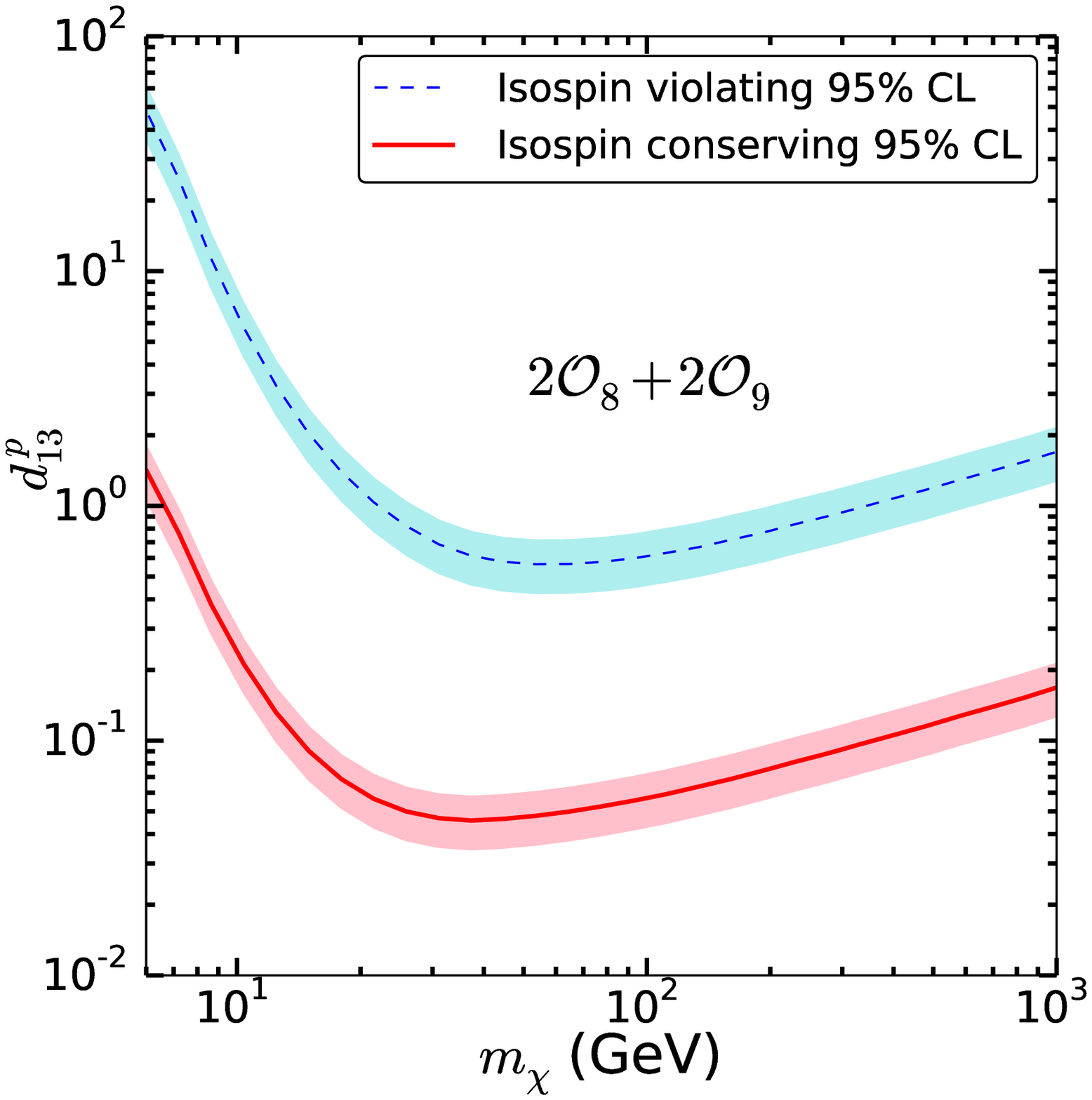}
\includegraphics[width=0.45\textwidth]{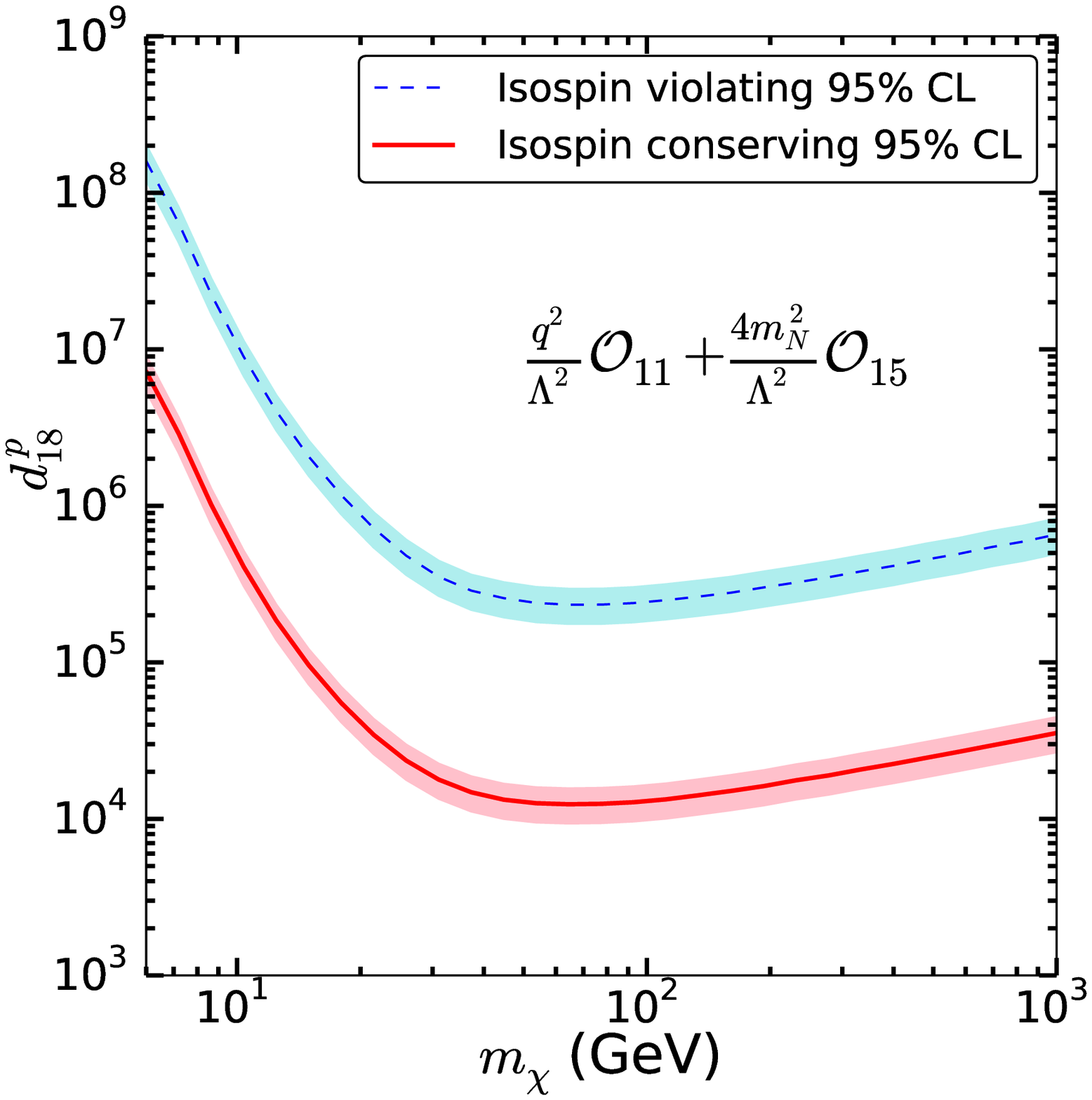}\\
\caption{The 95\% CL upper limits of all other couplings in 
Eq.~\eqref{eq:rlag} except $d_{12}$. 
%The red solid line is the ISC scenario and the blue dashed line is the 
%ISV scenario. The color bands demonstrate the astrophysical uncertainties 
%from the different approaches~\cite{Read:2014qva}. 
\label{fig:ds}}
\end{figure}

In previous sections, we have mentioned that all the four-point effective 
Lagrangians can be expanded by the NR effective operators shown in 
Table~\ref{tab:operators}. In fact, many relativistic Lagrangians 
correspond to one operator in the non-relativistic limit with a simple relationship (see Table 1 of 
Ref.~\cite{Anand:2013yka}). For example, the ratio of $d_3$ (the coupling 
of the interaction $i\bar{\chi}\gamma_5\chi\bar{N}N$) to $c_{11}$ (the 
coefficient of the operator $i{\bf{{S}}}_\chi\cdot\frac{{\bf{{q}}}}{m_N}$) 
is $-m_\chi/m_N$. Again here we use the notation following 
Ref.~\cite{Anand:2013yka}. 
However, for some relativistic Lagrangian $\mathcal{L}_i$ the expansions
of operators are not so trivial and may contain several operators, such as
\begin{eqnarray}
%d6
d_6\cdot
\bar{\chi} \gamma^\mu\chi \bar{N} i \sigma_{\mu \alpha} \frac{q^\alpha}{\Lambda} N
&\to &
d_6\cdot\left[
\frac{q^{2}}{ 2 m_N\Lambda}\mathcal{O}_1
-\frac{2m_N }{ \Lambda} \mathcal{O}_3
+\frac{2m_N^2}{m_\chi \Lambda}(\frac{q^2}{m_N^2}\mathcal{O}_4-\mathcal{O}_6)
\right],\nonumber \\
%7
d_7\cdot
\bar{\chi} \gamma^\mu\chi \bar{N} \gamma_{\mu}\gamma_5 N
&\to &
d_7\cdot\left[
-2\mathcal{O}_7
+\frac{2m_N }{ m_\chi} \mathcal{O}_9
\right],\nonumber \\
%9
d_9\cdot
\bar{\chi} i\sigma^{\mu\nu}\frac{q_\nu}{\Lambda}\chi \bar{N} \gamma_{\mu} N
&\to &
d_9\cdot\left[
-\frac{q^{2}}{ 2 m_N\Lambda}\mathcal{O}_1
+\frac{2m_N }{ \Lambda} \mathcal{O}_5
-\frac{2m_N}{\Lambda}(\frac{q^2}{m_N^2}\mathcal{O}_4-\mathcal{O}_6)
\right],\nonumber\\
%10
d_{10}\cdot
\bar{\chi} i \sigma^{\mu\nu} \frac{q_\nu}{\Lambda} \chi \bar{N} i \sigma_{\mu\alpha} 
\frac{q^\alpha}{\Lambda} N
&\to &
d_{10}\cdot\left[
\frac{4q^{2}}{\Lambda^2}\mathcal{O}_4
-\frac{4m_N^2}{\Lambda^2}\mathcal{O}_6
\right],\nonumber\\
%12
d_{12}\cdot
i \bar{\chi} i \sigma^{\mu\nu} \frac{q_\nu}{\Lambda} \chi 
\bar{N}i\sigma_{\mu\alpha}\frac{q^\alpha}{\Lambda}  \gamma_5 N
&\to &
d_{12}\cdot\left[
-\frac{m_N}{ m_\chi}\frac{q^{2}}{\Lambda^2}\mathcal{O}_{10}
-4\frac{q^2}{\Lambda^2}\mathcal{O}_{12} - 4 \frac{m_N^2}{\Lambda^2} \mathcal{O}_{15}
\right],\nonumber\\
%13 & $
d_{13}\cdot
\bar{\chi} \gamma^\mu \gamma_5 \chi \bar{N}\gamma_\mu N 
&\to &
d_{13}\cdot\left[
2\mathcal{O}_{8}+2\mathcal{O}_{9}
\right],\nonumber\\
%18 & $
d_{18}\cdot
i \bar{\chi} i\sigma^{\mu\nu}\frac{q_\nu}{\Lambda}\gamma_5 \chi \bar{N} i \sigma_{\mu \alpha} 
\frac{q^\alpha}{ \Lambda} N
&\to &
d_{18}\cdot\left[
\frac{q^2}{\Lambda^2}\mathcal{O}_{11}+\frac{4m_N^2}{\Lambda^2}\mathcal{O}_{15}
\right].
\label{eq:rlag}
\end{eqnarray}
Here we omitted the energy scale $\Lambda$ in front of the interactions.
It will be included in the computation as we did for the case of 
non-relativistic operators. 

%In this subsection, we will give the $95\%$ upper limit for $d_i$ where 
%$i=6,7,9,10,12,13,18$ which are formed by more than one operators. 
The effective Lagrangians $\mathcal{L}_9$ and $\mathcal{L}_{10}$ represent 
the magnetic dipole interaction with proton and neutron, respectively.    
The electric dipole moment interaction with proton is part of 
$\mathcal{L}_{18}$, 
and the Anapole interaction can be presented by 
$\mathcal{L}_{13}$. For more explicit expressions of the Anapole and 
electromagnetic moments in the effective theory, one can refer to
Ref.~\cite{Fitzpatrick:2012ib}.

To reuse the code developed for effective operators, we can use the 
relationship presented in Eq.~\eqref{eq:rlag}. Taking $\mathcal{L}_{13}$ 
as an example, if one sets its coupling as $d_{13}$, the coefficients of 
effective operators for the event rate computation are $c_8=2d_{13}$, 
$c_9=2d_{13}$, and the other coefficients are zero.

In the left panel of Fig.~\ref{fig:d12}, we present the upper limits on 
coupling $d^p_{12}$ for the ISC (red solid) and ISV (blue dashed) scenarios. 
For the ISV scenario for all Lagrangians in Eq.~\eqref{eq:rlag}, only 
the Lagrangian $\mathcal{L}_{12}$ has a positive maximum cancellation ratio.
Therefore we show its results separately from the others. The maximum 
cancellation ratio of neutron to proton couplings for $\mathcal{L}_{12}$ 
as a function of $\mchi$
is shown in the right panel.

\begin{figure}[!htb]
\includegraphics[width=0.45\textwidth]{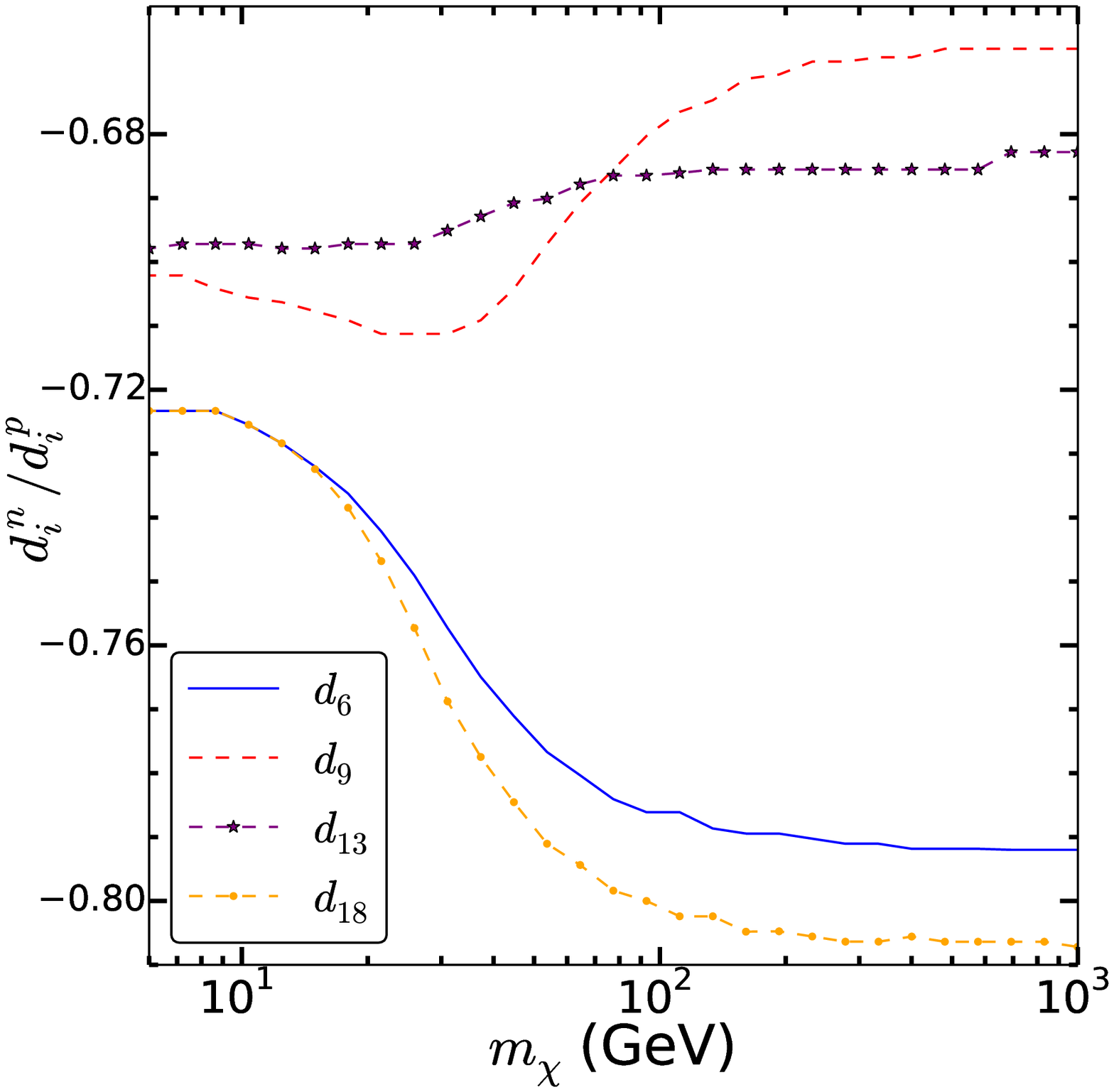}
\includegraphics[width=0.45\textwidth]{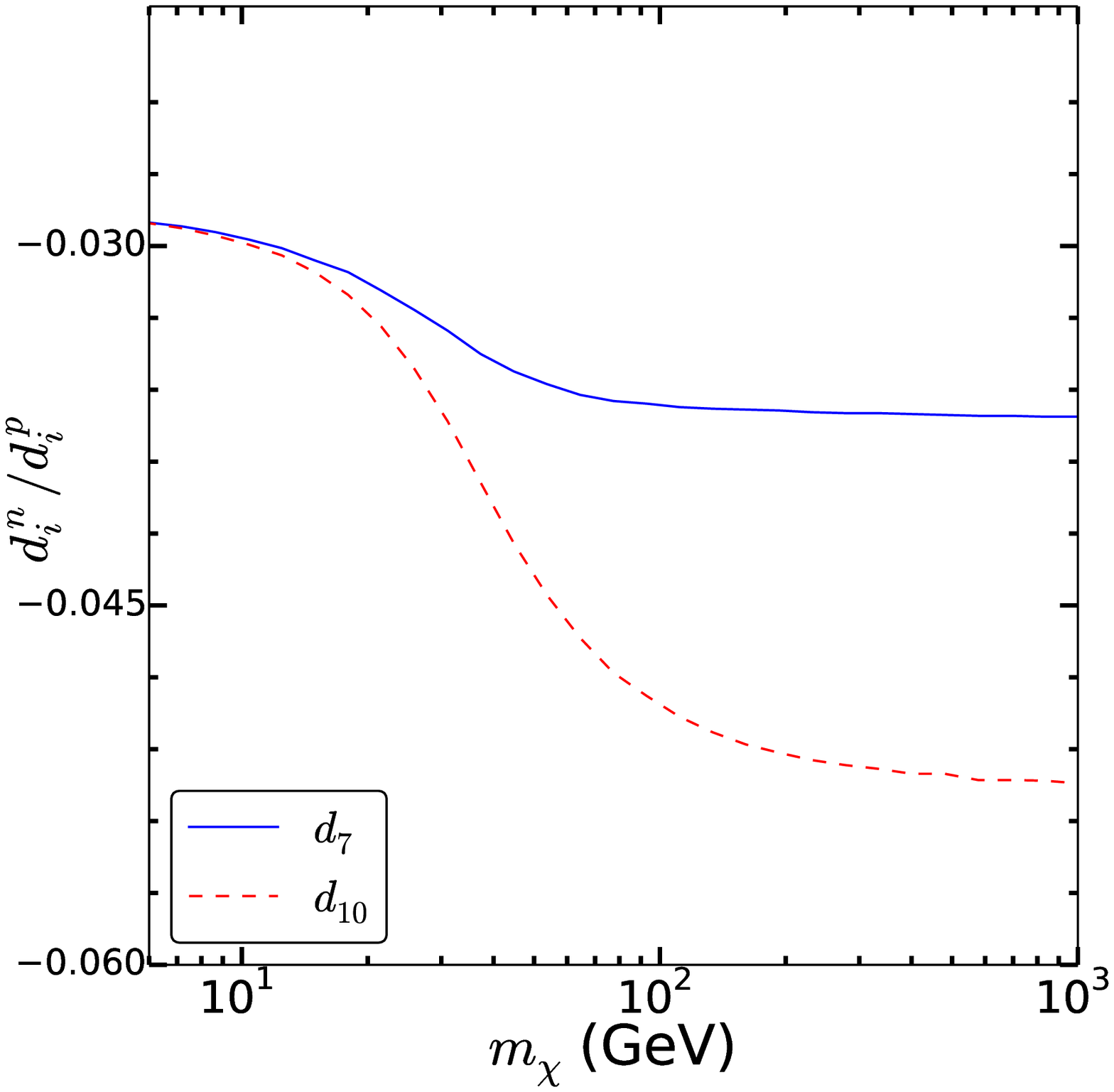}
\caption{The neutron to proton coupling ratios at the maximum cancellation 
for the effective Lagrangians ($i=6,7,9,10,13,18$).   
\label{fig:L_ISV}}
\end{figure}

The limits on the couplings for other Lagrangians are shown in 
Fig.~\ref{fig:ds}, and the maximum cancellation ratios of neutron to 
proton couplings are summarized in Fig.~\ref{fig:L_ISV}.

\subsection{High energy scale theory}

The new physics models beyond the standard model often appear at higher energy scale 
(i.e. greater than $Z$ boson mass),
 while the non-relativistic operators apply at the low 
energy scale. The match between these two scales may not be trivial.
For example, the long-distance corrections due to DM scattering 
with a pion exchanged between two nucleons can generate a coupling 
$c^i_{p,n}$ proportional to $q^{-2}$, which would significantly change 
the results~\cite{Fitzpatrick:2012ib,Bishara:2016hek}.
In this subsection, we adopt the \texttt{Mathematica} package 
\texttt{DirectDM}~\cite{Bishara:2016hek,Bishara:2017nnn}\footnote{We thank F.~Bishara, 
J.~Brod, B.~Grinstein and J.~Zupan for providing us the \texttt{DirectDM} 
code~\cite{Bishara:2016hek,Bishara:2017nnn}.} 
to calculate the relationship between the non-relativistic operators 
and the dimension-five and dimension-six effective DM-quark interactions. 
All contributions of quarks whose mass is less than $Z$ boson are included.

\begin{figure}[!htb]
\includegraphics[width=0.45\textwidth]{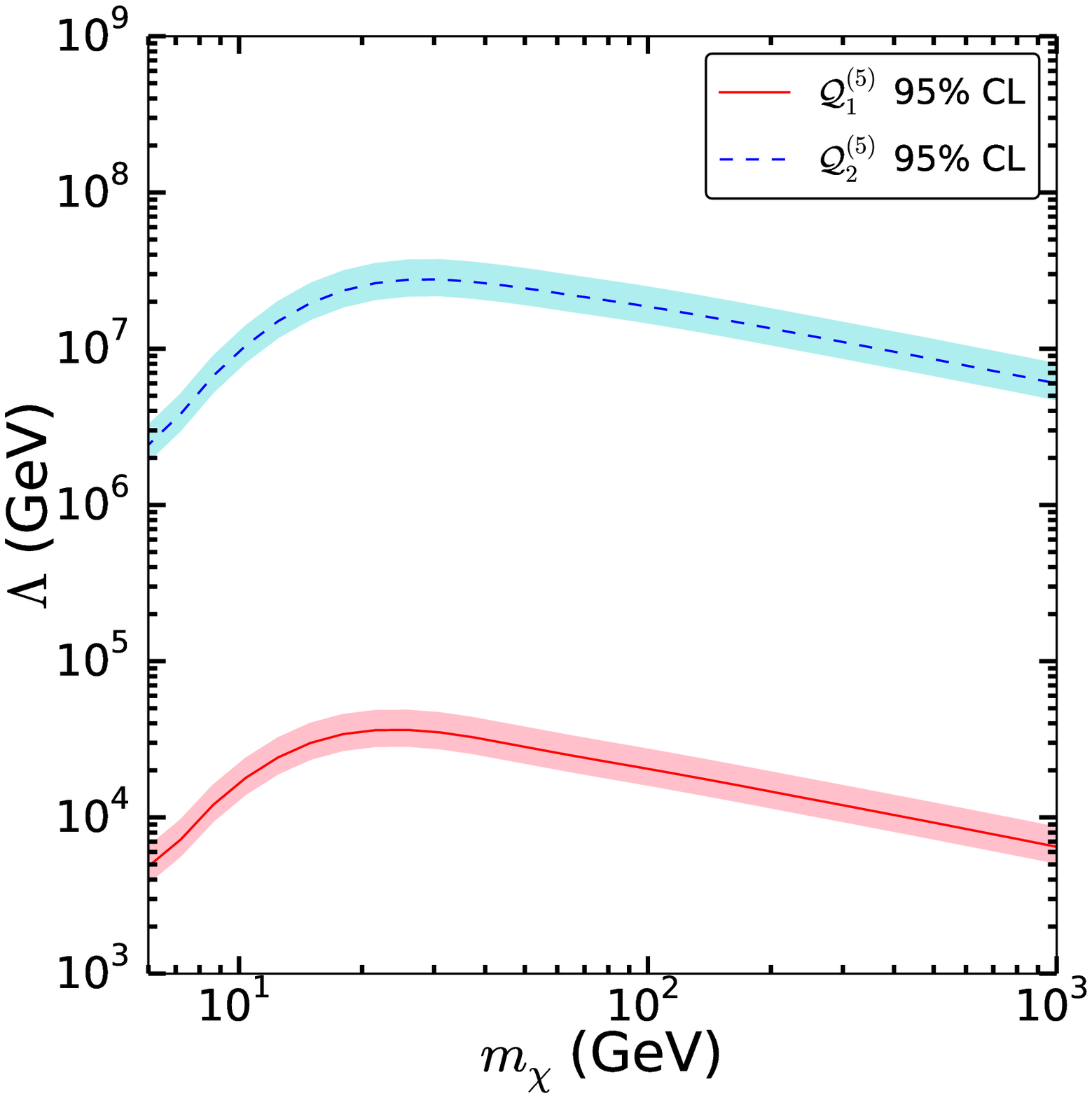}
\includegraphics[width=0.45\textwidth]{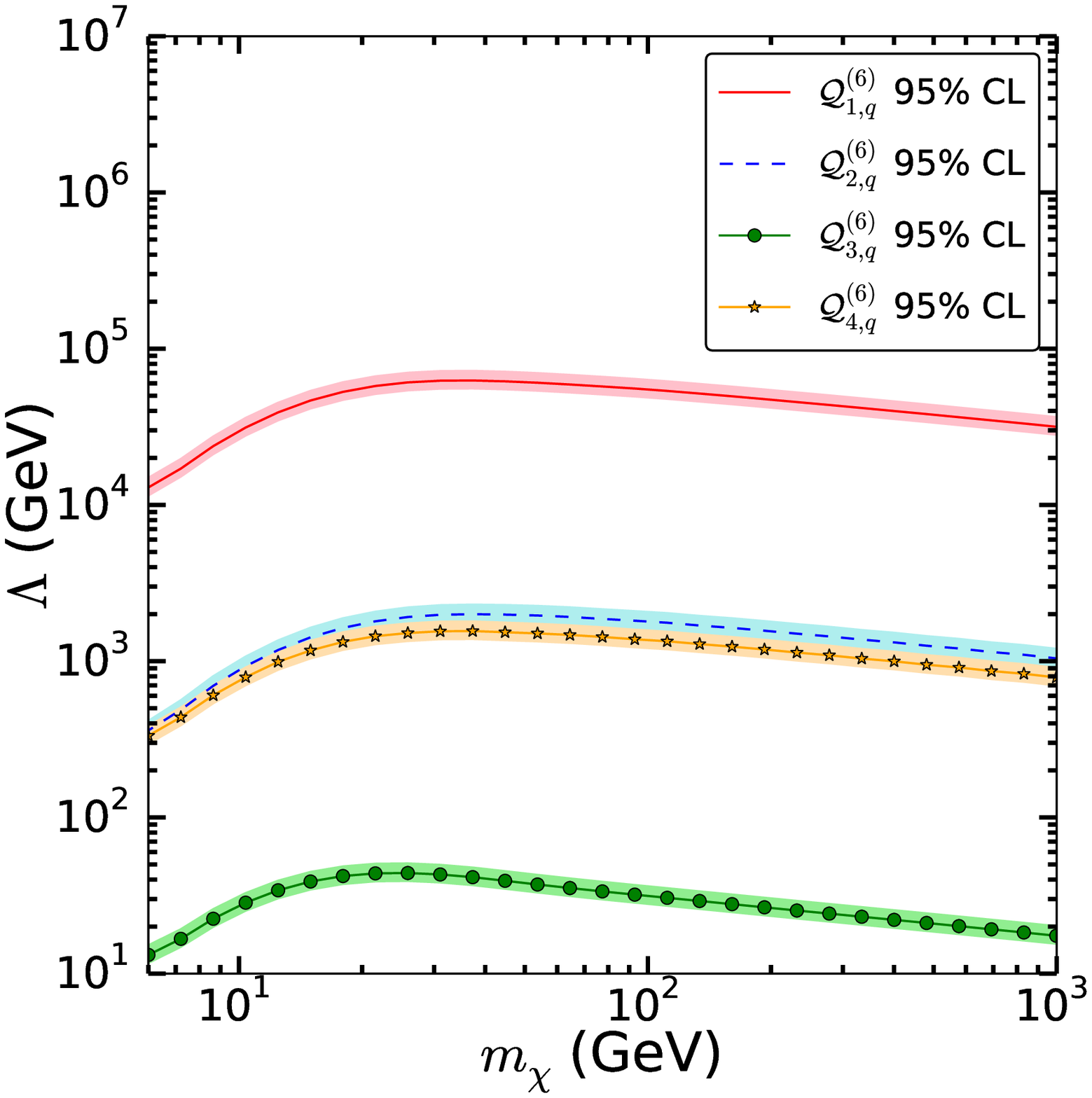}
\caption{The $95\%$ CL lower limits of the energy scale $\Lambda$ for 
dimension-five (left) and dimension-six (right) DM-quark interactions. 
The color bands demonstrate the uncertainties of the local DM density 
measurements~\cite{Read:2014qva}.
\label{fig:HEST}}
\end{figure}

Using similar conventions with that of Ref.~\cite{Bishara:2016hek}, 
we consider two dimension-five effective interactions,
\begin{equation}
\label{eq:dim5EW}
{\cal Q}_{1}^{(5)} = \frac{e}{8 \pi^2\Lambda} (\bar \chi \sigma^{\mu\nu}\chi)
 F_{\mu\nu} \,, 
 \qquad {\cal Q}_2^{(5)} = \frac{e }{8 \pi^2\Lambda} (\bar
\chi \sigma^{\mu\nu} i\gamma_5 \chi) F_{\mu\nu} \,,
\end{equation}
where $F_{\mu\nu}$ is the electromagnetic field strength tensor, and
$\Lambda$ is the new physics energy scale. The CP even operator ${\cal Q}_1^{(5)}$ 
and the CP odd operator ${\cal Q}_2^{(5)}$ represent the magnetic dipole 
and the electric dipole interactions, respectively. Because protons and 
neutrons couple to the electromagnetic field differently, and there is no 
electric dipole interaction between DM and neutrons, the isospin is not 
conserved in these dimension-five interactions. The interaction 
${\cal Q}_{1}^{(5)}$ can be expanded by operators with DM-proton 
coefficients $c^p_1,c^p_4,c^p_5,c^p_6$ and DM-neutron coefficients 
$c^n_4,c^n_6$, in which three coefficients $c^p_5,c^p_6,c^n_6$ are 
proportional to the inverse square of transfer momentum, $q^{-2}$. 
However, the interaction ${\cal Q}_{2}^{(5)}$ can only be expanded by 
operator $\mathcal{O}_{11}$ with the DM-proton coefficient $c^p_{11}$ 
which is also proportional to $q^{-2}$. See Appendix A of 
Ref.~\cite{Bishara:2016hek} for the exact relations.

In addition, we consider four dimension-six effective interactions between DM 
and quarks ($q$)
\begin{align}
{\cal Q}_{1,q}^{(6)} & = \frac{(\bar \chi \gamma_\mu \chi) (\bar q \gamma^\mu q)}{\Lambda^2},
 &{\cal Q}_{2,q}^{(6)} &= \frac{(\bar \chi\gamma_\mu\gamma_5 \chi)(\bar q \gamma^\mu q)}{\Lambda^2}, \nonumber
  \\ 
{\cal Q}_{3,q}^{(6)} & = \frac{(\bar \chi \gamma_\mu \chi)(\bar q \gamma^\mu \gamma_5 q)}{\Lambda^2}\,,
  & {\cal Q}_{4,q}^{(6)}& = \frac{(\bar
\chi\gamma_\mu\gamma_5 \chi)(\bar q \gamma^\mu \gamma_5 q)}{\Lambda^2}.\label{eq:dim6EW}
\end{align}
For the sake of simplicity, we only consider the scenario where
all the DM-quark couplings are unified. 
Therefore, isospin is conserved under such a simple assumption. 

Compared with dimension-five interactions, the mappings from these four 
interactions to the low energy DM-nucleon coefficients do not depend on 
the transfer momentum.  The interaction ${\cal Q}_{1,q}^{(6)}$ only 
relates to $c^{p,n}_1$. 
In addition to $c^{p,n}_1$,
${\cal Q}_{2,q}^{(6)}$ contains two more coefficients $c^{p,n}_8$ and 
$c^{p,n}_9$. The interaction ${\cal Q}_{3,q}^{(6)}$ has one more relevant 
coefficient $c^{p,n}_7$ added to that of ${\cal Q}_{2,q}^{(6)}$, and the 
interaction ${\cal Q}_{4,q}^{(6)}$ has one additional relevant coefficient 
$c^{p,n}_4$ added to that of ${\cal Q}_{3,q}^{(6)}$.

In Fig.~\ref{fig:HEST}, we present the $95\%$ lower limits on $\Lambda$ 
for dimension-five interactions (left) and dimension-six interactions (right). 
It is shown that the constaints on the electric dipole interaction ${\cal Q}_{2}^{(5)}$ 
are stronger than that of the magnetic dipole interaction ${\cal Q}_{1}^{(5)}$.
The dominant 
operators among the six operators in ${\cal Q}_{1}^{(5)}$  NR expansion do not depend on $q^2$; 
${\cal Q}_{2}^{(5)}$ only relates to 
$\mathcal{O}_{11}$ which is inversely proportional to $q^2$. 
In the right panel of Fig.~\ref{fig:HEST}, 
the interaction ${\cal Q}_{4,q}^{(6)}$ has a significant positive contribution 
from $\mathcal{O}_4$ which makes the energy scale of ${\cal Q}_{4,q}^{(6)}$ 
larger than that of ${\cal Q}_{3,q}^{(6)}$.

\section{Summary\label{sec:sum}}

As the first time to consider a combined analysis of three most 
recent and powerful experimental data sets from PandaX, LUX, and XENON1T, 
we have approximately reconstructed their likelihood in terms of event 
rates and reported a new combined limit based on a spin-1/2 DM. 
To consider the possible impact of astrophysical uncertainties from the DM local velocity, 
escape velocity, and local density, we introduce these parameters as nuisance 
parameters in our likelihood. 

In the low energy effective operator framework, we apply our combined 
likelihood for each effective operator and derive their $95\%$ CL upper 
limits on the plane ($m_\chi$, $c_i^p$) for the ISC and ISV (with maximum cancellations) scenarios. 
As expected, our combined limits of the effective operators are more 
stringent than previous studies~\cite{Fitzpatrick:2012ib,Catena:2015uua}.
In addition to the low energy effective operator 
framework, we also study the high energy effective Lagrangians which can 
usually be expressed as the combination of several operators. 
The $95\%$ CL limits (upper or lower) on the ($m_\chi$, $d_i^p$) and 
($m_\chi$, $\Lambda$) planes are presented with several representative 
high energy effective Lagrangians. 

The inclusion of the uncertainties of the DM velocity distribution
and the escape velocity of the solar location leads to $\lesssim 5\%$
uncertainties for the velocity independent operators and $\lesssim 7.5\%$ 
for the velocity dependent operators. The maximum uncertainties appear
around $m_\chi\lesssim 10\gev$ whose recoil energies 
fall into the small efficiency region. However, such 
astrophysical parameters depend on the data sets and modeling of the 
luminous matter distribution, as well as the prior assumption of the
DM density profile, which may be subject to additional systematic
uncertainties. Therefore we quote a relatively large band of the
local DM density measured by different analyses~\cite{Read:2014qva}
to show a potential uncertainty range of our results.

For the xenon target detectors, we report new ISV coupling ratios $c_i^n/c_i^p$ 
for the maximum cancellation between the 
contributions of DM-proton and DM-neutron couplings. Firstly, we find that 
the ratio is not a constant with respect to the DM mass. However, as long 
as the DM mass is heavy, the ratio asymptotically
approaches a constant. Secondly, only the operator $\mathcal{O}_{13}$ has
a positive ratio, and the rest operators have negative ratios for the 
maximum cancellation. Finally, the well-known number of $c_i^n/c_i^p=-0.7$ 
only agrees with the operators $\mathcal{O}_{i=1,3,5,8,11,12,15}$. 
For the high energy effective Lagrangian cases, only the transfer 
momentum dependent operators can have a large ratio change with respect 
to the DM mass. 
%The ratios are between those of their compositions, which are lower energy operators. 
The ratios are between their compositions of lower energy operators.
In this paper, we only studied one-nucleon contributions.  
If the two-nucleon correlations would be included, 
the result of ISV case can be changed~\cite{Cirigliano:2013zta,Korber:2017ery}. 

Lastly, we would like to suggest the future experiments to 
publish their limits of event rates $\mathcal{R}$ togehter with the 
spin-independent and spin-dependent cross sections. There are two 
advantages to present the limits on the plane ($m_\chi$, $\mathcal{R}$).  
Firstly, this can help theorists to go beyond the neutralino-like 
benchmark scenario which is only relevant to operators $\mathcal{O}_1$ 
and $\mathcal{O}_4$. As performed by the SuperCDMS 
collaboration~\cite{Schneck:2015eqa} and XENON collaboration~\cite{Aprile:2017aas}, 
the EFT operators can provide a more general framework to explore the DM-nuclear interaction. 
With the likelihood information of $\mathcal{R}$, one can simply obtain 
the experimental limits for their interested operators. Secondly, it 
also helps to unfold the astrophysical uncertainties. The systematic 
uncertainties for the next generation DM direct detection detectors 
will be more important because the statistical precision can 
be much improved for the future ton-scale detectors and more kinematic 
information can be obtained for example in the DM directional detection experiments.

\appendix
%%%%%%%%%%%%%%%%%%%%
\section{DM response function}
\label{sec:DM_respond}
%%%%%%%%%%%%%%%%%%%%

%In this section, we list all the DM response functions. 
Following the definition of Ref.~\cite{Anand:2013yka}, we use the 
notations $M$, $\Delta$, $\Sigma^\prime$, $\Sigma^{\prime\prime}$, 
$\tilde{\Phi}^\prime$, and $\Phi^{\prime\prime}$ to represent the DM 
currents by the vector charge, vector transverse magnetic, axial 
transverse electric, axial longitudinal, vector transverse electric, 
and vector longitudinal operators, respectively. In the DM response 
functions, one needs also to consider two interference terms, 
$\Phi^{\prime\prime}M$ and $\Delta\Sigma^\prime$. There are eight 
possible DM response functions: 
\begin{eqnarray}
%vector charge
 R_{M}^{\tau \tau^\prime}\left(v_T^{\perp 2}, {q^2 \over m_N^2}\right) &=& 
 c_1^\tau c_1^{\tau^\prime } + {j_\chi (j_\chi+1) \over 3} \left[ {q^2 \over m_N^2} v_T^{\perp 2} c_5^\tau c_5^{\tau^\prime }+v_T^{\perp 2}c_8^\tau c_8^{\tau^\prime }
+ {q^2 \over m_N^2} c_{11}^\tau c_{11}^{\tau^\prime } \right] \nonumber \\
%vector transverse magnetic
 R_{\Phi^{\prime \prime}}^{\tau \tau^\prime}\left(v_T^{\perp 2}, {q^2 \over m_N^2}\right) &=& 
 {q^2 \over 4 m_N^2} c_3^\tau c_3^{\tau^\prime } + {j_\chi (j_\chi+1) \over 12} \left( c_{12}^\tau-{q^2 \over m_N^2} c_{15}^\tau\right) \left( c_{12}^{\tau^\prime }-{q^2 \over m_N^2}c_{15}^{\tau^\prime} \right)  \nonumber \\
 R_{\Phi^{\prime \prime} M}^{\tau \tau^\prime}\left(v_T^{\perp 2}, {q^2 \over m_N^2}\right) &=&  c_3^\tau c_1^{\tau^\prime } + {j_\chi (j_\chi+1) \over 3} \left( c_{12}^\tau -{q^2 \over m_N^2} c_{15}^\tau \right) c_{11}^{\tau^\prime } \nonumber \\
  R_{\tilde{\Phi}^\prime}^{\tau \tau^\prime}\left(v_T^{\perp 2}, {q^2 \over m_N^2}\right) &=&
  {j_\chi (j_\chi+1) \over 12} \left[ c_{12}^\tau c_{12}^{\tau^\prime }+{q^2 \over m_N^2}  c_{13}^\tau c_{13}^{\tau^\prime}  \right] \nonumber \\
   R_{\Sigma^{\prime \prime}}^{\tau \tau^\prime}\left(v_T^{\perp 2}, {q^2 \over m_N^2}\right)  &=&
   {q^2 \over 4 m_N^2} c_{10}^\tau  c_{10}^{\tau^\prime } +
  {j_\chi (j_\chi+1) \over 12} \left[ c_4^\tau c_4^{\tau^\prime} + \right.  \nonumber \\
 && \left. {q^2 \over m_N^2} ( c_4^\tau c_6^{\tau^\prime }+c_6^\tau c_4^{\tau^\prime })+
 {q^4 \over m_N^4} c_{6}^\tau c_{6}^{\tau^\prime } +v_T^{\perp 2} c_{12}^\tau c_{12}^{\tau^\prime }+{q^2 \over m_N^2} v_T^{\perp 2} c_{13}^\tau c_{13}^{\tau^\prime } \right] \nonumber \\
    R_{\Sigma^\prime}^{\tau \tau^\prime}\left(v_T^{\perp 2}, {q^2 \over m_N^2}\right)  &=&{1 \over 8} \left[ {q^2 \over  m_N^2}  v_T^{\perp 2} c_{3}^\tau  c_{3}^{\tau^\prime } + v_T^{\perp 2}  c_{7}^\tau  c_{7}^{\tau^\prime }  \right]
       + {j_\chi (j_\chi+1) \over 12} \left[ c_4^\tau c_4^{\tau^\prime} +  \right.\nonumber \\
       &&\left. {q^2 \over m_N^2} c_9^\tau c_9^{\tau^\prime }+{v_T^{\perp 2} \over 2} \left(c_{12}^\tau-{q^2 \over m_N^2}c_{15}^\tau \right) \left( c_{12}^{\tau^\prime }-{q^2 \over m_N^2}c_{15}^{\tau \prime} \right) +{q^2 \over 2 m_N^2} v_T^{\perp 2}  c_{14}^\tau c_{14}^{\tau^\prime } \right] \nonumber \\
% vector transverse magnetic
     R_{\Delta}^{\tau \tau^\prime}\left(v_T^{\perp 2}, {q^2 \over m_N^2}\right)&=&  {j_\chi (j_\chi+1) \over 3} \left[ {q^2 \over m_N^2} c_{5}^\tau c_{5}^{\tau^\prime }+ c_{8}^\tau c_{8}^{\tau^\prime } \right] \nonumber \\
 R_{\Delta \Sigma^\prime}^{\tau \tau^\prime}\left(v_T^{\perp 2}, {q^2 \over m_N^2}\right)&=& {j_\chi (j_\chi+1) \over 3} \left[c_{5}^\tau c_{4}^{\tau^\prime }-c_8^\tau c_9^{\tau^\prime} \right].
\label{eq:Response}
\end{eqnarray}

%%%%%%%%%%%%%%%%%%%%
\section{Integration of the DM velocity distribution over solid angle}
\label{sec:DM_vf}
%%%%%%%%%%%%%%%%%%%%
From the differential event rate Eq.~\eqref{Eq:dndQ}, one can compute 
the solid angle integration part first, namely  
%$$\frac{{\rm d}\mathcal{R}}{{\rm d}Q} = \sum_{T} \xi_T \frac{\rho_{0}}{m_\chi m_{T} }  
% \int_{v > v_{\rm min}(Q)} \,  v  f(\vec{v} + \vec{v}_e)\frac{{\rm d}\sigma}{{\rm d}Q}\, d^3v,$$
%$$ first. 
\begin{equation}
%\int_{v > v_{\rm min}(Q)} \,  v  f(\vec{v} + \vec{v}_e)\frac{{\rm d}\sigma}{{\rm d}Q}\, d^3v= 
\int_{v > v_{\rm min}(Q)} v^{3} d v \int_{0}^{2 \pi} 
d \varphi \int_{0}^{\pi} \frac{{\rm d}\sigma}{{\rm d}Q} f(\vec{v} + 
\vec{v}_e) \sin \theta d\theta, 
\label{eq:dndq2}
\end{equation}
where $\varphi$ and $\theta$ are the azimuthal angle and polar angle in 
the Earth frame. Because $\frac{{\rm d}\sigma}{{\rm d}Q}$ 
does not depend on $\theta$ and $\varphi$, we can rewrite  
Eq.~\eqref{eq:dndq2} as 
\begin{equation}
\int_{v > v_{\rm min}(Q)} v^{3} \frac{{\rm d}\sigma}{{\rm d}Q} \times I(\vec{v} + \vec{v}_e) d v,  
%\int_{0}^{2 \pi} d \varphi \int_{0}^{\pi} f(\vec{v} + \vec{v}_e) \sin \theta d \theta 
\label{eq:dndq3}
\end{equation}
where
\begin{equation}
I(\vec{v} + \vec{v}_e)\equiv\int_{0}^{2 \pi} d \varphi \int_{0}^{\pi} f(\vec{v} + 
\vec{v}_e) \sin \theta d \theta. 
\label{eq:I} 
\end{equation}      
%and then we will demonstrate the integration of $I$. 
  
We adopt the soft truncated Maxwell-Boltzmann distribution of $f$:
\begin{equation}
f(\vec{v} + \vec{v}_e)=\frac{(e^{-{(\vec{v} + \vec{v}_e)^{2}}/{v_{0}^{2}}}-e^{-{v_\text{esc}^{2}}/{v_{0}^{2}}})\Theta(v_\text{esc}^{2}-(\vec{v} + \vec{v}_e)^{2})}
{\pi^{3/2} v_{0}^{3}\times\rm{Norm}},
\label{eq:I2} 
\end{equation}  
where $\Theta(x)$ is the step function, and ``Norm'' is the normalization 
factor
\begin{equation*}
{\rm erf}\left(\frac{v_\text{esc}}{v_{0}}\right)-
\left[\frac{4}{3\sqrt{\pi}} \left(\frac{v_\text{esc}}{v_{0}}\right) ^{3} 
+  \frac{2v_\text{esc}}{\sqrt{\pi}v_{0}}\right]
\exp\left(-\frac{v_\text{esc}^{2}}{v_{0}^{2}}\right).
\end{equation*}  
In the above formulae, $v_{0}$, $v_{e}$ and $v_\text{esc}$ are all in the
Galactic center (GC) frame. 

The relationship between the DM velocity $\vec{v} + \vec{v}_e$ in the GC 
frame and the velocity magnitude $v$ in the Earth frame is
\begin{equation*}
\vec{v} + \vec{v}_e=(v\sin\theta\cos\varphi ,
                     v\sin\theta\sin\varphi ,
                     v\cos \theta + v_{e}).
\end{equation*}  
Note that we set $\vec{v_{z}}$ to be parallel to $\vec{v_{e}}$. Hence, 
%$$(\vec{v} + \vec{v}_e)^{2}=v^{2}+v_{e}^{2}+2 v v_{e} cos \theta .$$
the integration of angle $\varphi$ in Eq.~\eqref{eq:I} can be done prior 
to $\theta$, which gives a prefactor $2\pi$.   
% $$I = 2 \pi \int_{0}^{\pi} f(\vec{v} + \vec{v}_e) sin \theta  d \theta.$$
Finally, we can integrate $I(\vec{v} + \vec{v}_e)$ over the angle $\theta$ 
in terms of variables $ a = v_\text{esc}/v_{0} $, $ b = v_{e}/v_{0} $, and 
$ x = v/v_{0} $.   
%        So 
%        $$norm= erf(a)-\frac{4}{3} \pi^{-1/2} a^3 e^{- a^2}- 2 \pi^{-1/2} a e^{- a^2}.$$
We find that $I(\vec{v} + \vec{v}_e)$ has 3 types of solutions, depending on
the relationships between $a$, $b$ and $x$.
\begin{enumerate}
\item Region $x+b<a$ or $v+v_{e}<v_{esc}$: 
        $$I=\frac{e^{-(x+b)^{2}} \left[3  e^{a^2} \left(e^{4 x b}-1 \right)
        -12 x b e^{(x+b)^{2}}\right]}{x b v_{0}^3 \left[3 \sqrt{\pi } e^{a^2} 
        {\rm erf} \left(a \right)-6 a-4 a^3 \right]}.$$
\item Region ($x+b>a$, $|x-b|<a$) or ($v+v_{e}>v_{esc}$, $|v-v_{e}|<v_{esc}$): 
$$I=\frac{3 \left[e^{a^2-(x-b)^2}+(x-b)^2-a^2-1\right]}{x b v_{0}^3 \left[3 \sqrt{\pi } e^{a^2} 
{\rm erf}\left(a \right)-6 a -4 a^3 \right]}.$$
\item Region $|x-b|>a$ or $|v-v_{e}|>v_{esc}$:        
        $$I = 0.$$
\end{enumerate}

%
%        1.$x+b<a$ ($v+v_{e}<v_{esc}$)
%        $$I = 2 \pi \int_{0}^{\pi} \frac{1}{\pi^{3/2} v_{0}^{3} norm} (e^{-(x^{2}+b^{2}+2 x b cos \theta)}-e^{-a^2}) sin \theta  d \theta,$$
%        $$I=\frac{e^{-(x+b)^{2}} \left(3  e^{a^2} \left(e^{4 x b}-1 \right)-12 x b e^{(x+b)^{2}}\right)}{x b v_{0}^3 \left(3 \sqrt{\pi } e^{a^2} erf \left(a \right)-6 a-4 a^3 \right)}.$$
%        \\
%        2.$x+b>a,|x-b|<a$ ($v+v_{e}>v_{esc},|v-v_{e}|<v_{esc}$)
%        $$I = 2 \pi \int_{arccos \frac{a^{2}-x^{2}-b^{2}}{2 x b}}^{\pi} \frac{1}{\pi^{3/2} v_{0}^{3} norm} (e^{-(x^{2}+b^{2}+2 x b cos \theta)}-e^{-a^2}) sin \theta  d \theta,$$
%        $$I=\frac{3 \left( \left(e^{a^2-(x-b)^2}-1\right)+(x-b)^2-a^2\right)}{x b v_{0}^3 \left(3 \sqrt{\pi } e^{a^2} erf\left(a \right)-6 a -4 a^3 \right)}.$$
%        \\
%        3.$|x-b|>a$ ($|v-v_{e}|>v_{esc}$)        
%         $$I = 0.$$

\section{Confidence limits}\label{sec:CLb}
To determine the exclusion limits, a confidence level ($CL$) is usually 
used, defined as  
\begin{equation}
1-CL=\int_{\chi^2_c}^\infty \mathcal{L}(\chi^2) d\chi^2,
\label{eq:CL} 
\end{equation} 
where the desired $CL$ value is the integral likelihood for the range 
$0<\chi^2<\chi^2_c$. If the likelihood $\mathcal{L}$ is Gaussian 
distribution ($\propto\exp(-\chi^2/2)$), then we can simply obtain 
$\chi^2_c=2.71$ at $CL=0.95$ (one sided confidence limit). Using 
such a definition, conventionally we can describe the consistency of 
the data with the \textit{background-only hypothesis} $CL_b$ and the 
\textit{signal-plus-background hypothesis} $CL_{s+b}$.

However, in many real cases one can introduce nuisance parameters and 
model parameters to the likelihood function as systematic uncertainties 
which makes it difficult to find $CL_b$ and $CL_{s+b}$ by integrating 
the multi-dimensional parameter space in Eq.~\eqref{eq:CL}. Therefore, 
it is convenient to introduce a test-statistic 
$\chi^2(s,b)=-2\ln\mathcal{L}(s,b)$ to numerically 
%count the $CL_{s+b}$ and $CL_b$, namely,    
pin down the confidence level for two hypotheses 
of \texttt{background-only} and \texttt{signal-plus-background.}

In the $m_{\chi}-\mathcal{R}$ plane, we divide $m_{\chi}$ into various 
bins. For each $m_{\chi}$ bin,  
%the $\Delta m_{\chi}$ is so small that in each bin 
the $\chi^2$ varies as a function of $\mathcal{R}$.  
Then, we solve the following equation
\begin{equation}
\Delta \chi^2=\chi^2(m_{\chi},\mathcal{R})-\chi^2(m_{\chi},\mathcal{R}=0)=2.71,
\end{equation}
to get a value of $\mathcal{R}_{95}$ corresponding to the 95\% upper limits  
based on the $CL_b$ method. Repeating the process for different $m_{\chi}$ 
bins, we can get a curve on the $m_{\chi}-\mathcal{R}_{95}$ plane 
corresponding to the 95\% exclusion limit.

Regarding to $CL_{s+b}$ method, the same procedure as described above 
can be performed, but solving a different equation
\begin{equation}
\Delta \chi^2=\chi^2(m_{\chi},\mathcal{R})-\chi^2_{\rm min}(m_{\chi},\mathcal{R})=2.71
\end{equation}
to get the $\mathcal{R}_{95}$ limits.

Since we are using the null hypothesis (background only), we adopt the
$CL_b$ method to evaluate the limits. It is worthy mentioning that our 
limits will be hence slightly weaker than the $CL_{s+b}$ method.

\section{Limit including PandaX run10 data}\label{sec:PX10}

\begin{figure}[!htb]
\includegraphics[width=0.45\textwidth]{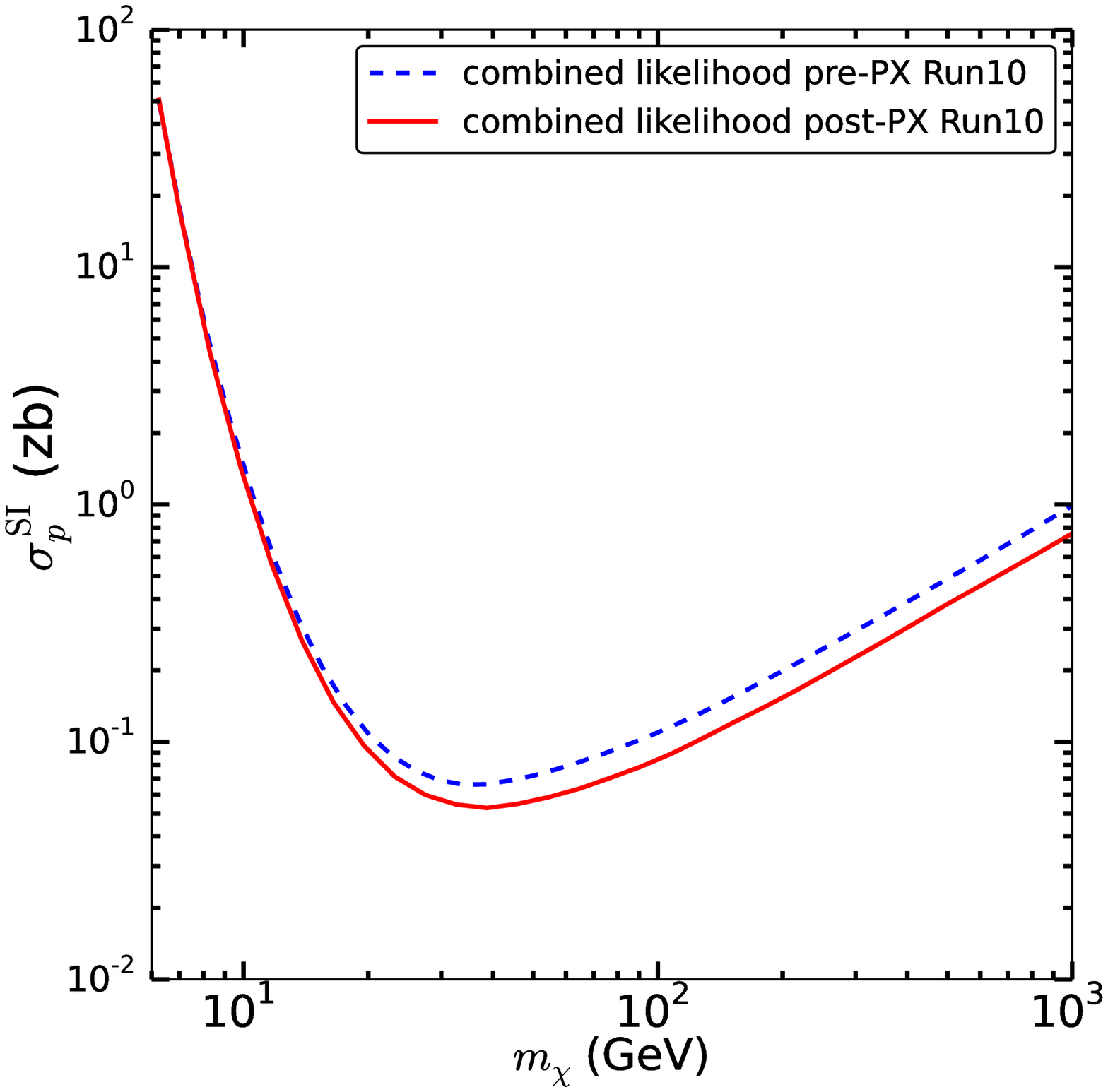}
\includegraphics[width=0.45\textwidth]{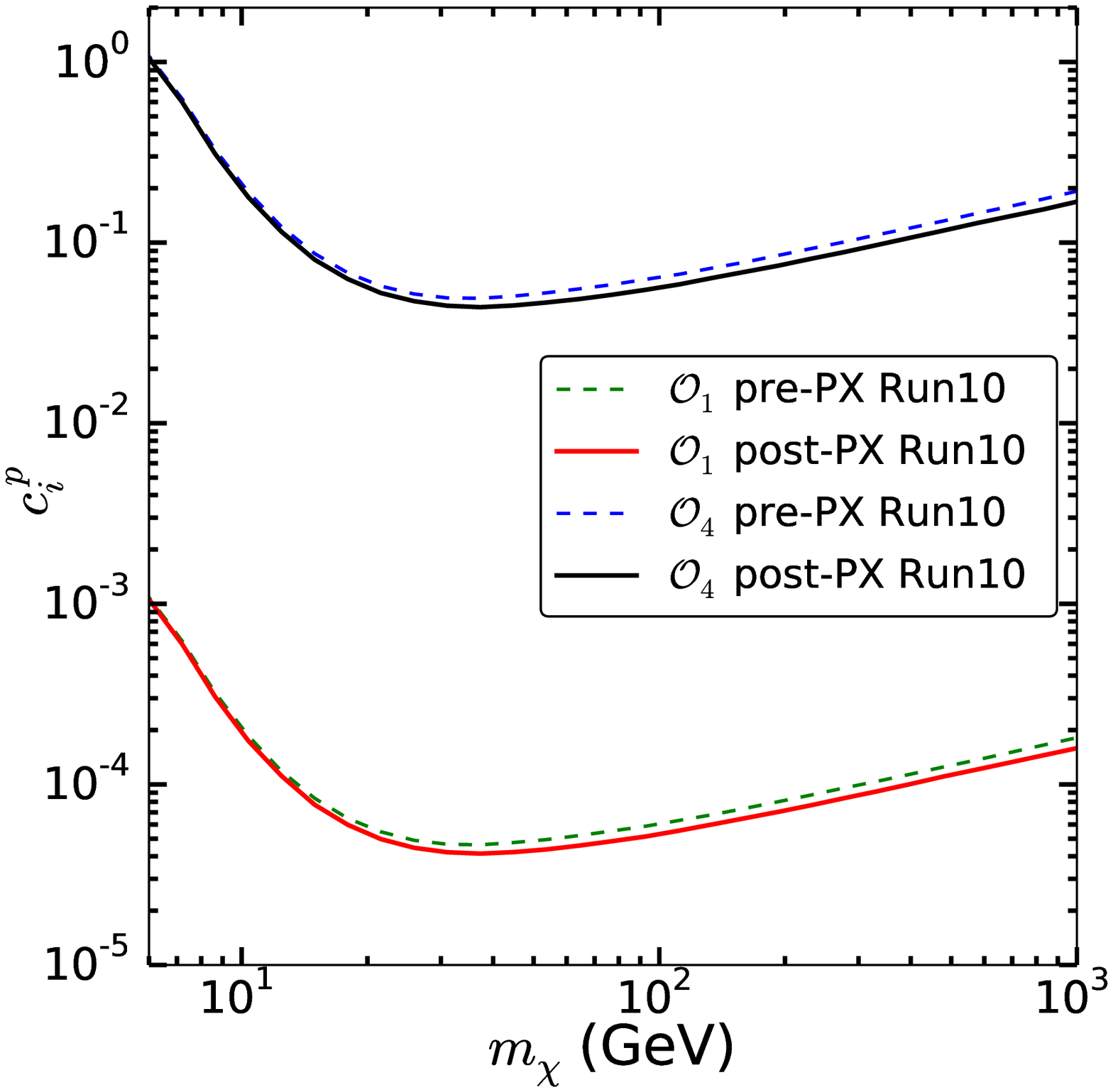}
\caption{Left panel: the $95\%$ CL upper limits of $\sigsip$ 
for new combined likelihood (red solid line) and old combined likelihood (blue dashed line). 
Right panel: the $95\%$ CL upper limits of $c_i^p$ for $\mathcal{O}_1$ and $\mathcal{O}_4$ (ISC). 
In both panels, the solid lines are for the old combined likelihood but dash lines are the new one.  
\label{fig:PX10}}
\end{figure}

In this appendix, we will show the update limit including the latest 
PandaX \texttt{run9$'$+run10} data~\cite{Cui:2017nnn}. 
For the new \texttt{Run9$'$}, the total observed event number after all the cuts is 1 and the
expected background event number is $3.2\pm 0.9$. 
For the \texttt{Run10}, there is none observed event number
but the expected background event number is $1.8\pm0.5$.
With the updated exposure ${\cal E}_9=26180.44$ kg-day 
and ${\cal E}_{10}=27871.65$ kg-day, we replace the old PandaX likelihood (\texttt{run8+run9}) 
to new one (\texttt{run9$'$+run10}) in the Fig.~\ref{fig:PX10}. 
One can see that the improvement is around $20\%$ in the spin independent cross section but 
it is not significant in the ($m_\chi$,~$c_i^p$) plane.

\section*{Acknowledgments}
We would like to thank Joachim Brod and Jianglai Liu for useful discussion. 
The work of Z.L.\ is supported in part by the Nanjing University Grant 14902303.
Y.S.T. would like to thank the hospitality of Technische Universit\"at 
Dortmund where part the work was completed. Q.Y. acknowledges the support 
by the 100 Talents program of Chinese Academy of Sciences.

%%%%%%%%%%%%%%%%%%%%%%%%%%%%%%%%%%%%%%%%%%%%%%%%%%%%

\bibliographystyle{unsrt}

\end{document}